\documentclass[isre,dblanonrev]{informs4_new} 

\usepackage{arydshln}
\usepackage{multirow}
\usepackage{enumitem}
\usepackage{amsmath}
\usepackage{graphicx}

\usepackage{url}
\usepackage{cases}
\usepackage{mathtools}
\usepackage{booktabs}
\usepackage{hyperref}
\hypersetup{
    colorlinks=true,
    urlcolor={blue},
    citecolor=[rgb]{0,0.5,0.5},
}

\usepackage{algorithm}
\usepackage{algorithmic}
\usepackage{newfloat}
\usepackage{listings}
\floatstyle{ruled}
\newfloat{listing}{tb}{lst}{}
\floatname{listing}{Listing}
\usepackage{booktabs,multirow,diagbox,enumitem}
\usepackage{amsfonts,nicefrac}
\usepackage{multirow,tabularx,diagbox}
\usepackage{microtype}      
\usepackage{dblfloatfix}     
\usepackage{cuted}
\usepackage[switch]{lineno} 
\usepackage{rotating} 
\usepackage{arydshln}
\usepackage{makecell,enumitem}
\usepackage[xcdraw]{xcolor}
\usepackage[font=scriptsize]{subfig}
\usepackage{graphics}
\usepackage{booktabs}
\usepackage{amsmath}
\usepackage{MnSymbol}
\usepackage{wasysym}
\usepackage{eurosym}
\usepackage{adjustbox}

\usepackage{url}
\makeatletter
\g@addto@macro{\UrlBreaks}{\UrlOrds}
\makeatother

\newcommand{\vpara}[1]{\vspace{0.05in}\noindent\textbf{#1 }}
\newcommand{\hide}[1]{} 

\renewcommand{\ell}{\mathcal L}

\DoubleSpacedXI 

\usepackage{natbib}
 \bibpunct[, ]{(}{)}{,}{a}{}{,}%
 \def\bibfont{\small}%
\TheoremsNumberedThrough     

\EquationsNumberedThrough    

\begin{document}

\setcounter{page}{1}

\RUNTITLE{Secure Confidential Business Information When Sharing Machine Learning Models}

\TITLE{Secure Confidential Business Information When Sharing Machine Learning Models}

\ARTICLEAUTHORS{%
\AUTHOR{Yunfan Yang}
\AFF{School of Management, Fudan University, \EMAIL{yunfanyang24@m.fudan.edu.cn}}
\AUTHOR{Jiarong Xu}
\AFF{School of Management, Fudan University, \EMAIL{jiarongxu@fudan.edu.cn}}
\AUTHOR{Hongzhe Zhang}
\AFF{School of Management and Economics, The Chinese University of Hong Kong, Shenzhen, \EMAIL{zhanghongzhe@cuhk.edu.cn}}
\AUTHOR{Xiao Fang}
\AFF{Lerner College of Business and Economics, University of Delaware, \EMAIL{xfang@udel.edu}}
} 

\OneAndAHalfSpacedXI
\ABSTRACT{
Model-sharing offers significant business value by enabling firms with well-established Machine Learning (ML) models to monetize and share their models with others who lack the resources to develop ML models from scratch. However, concerns over data confidentiality remain a significant barrier to model-sharing adoption, as Confidential Property Inference (CPI) attacks can exploit shared ML models to uncover confidential properties of the model provider's private model training data. Existing defenses often assume that CPI attacks are non-adaptive to the specific ML model they are targeting. This assumption overlooks a key characteristic of real-world adversaries: their responsiveness, i.e., adversaries' ability to dynamically adjust their attack models based on the information of the target and its defenses. 
To overcome this limitation, we propose a novel defense method that explicitly accounts for the responsive nature of real-world adversaries via two methodological innovations: a novel Responsive CPI attack and an attack–defense arms race framework. The former emulates the responsive behaviors of adversaries in the real world, and the latter iteratively enhances both the target and attack models, ultimately producing a secure ML model that is robust against responsive CPI attacks. Furthermore, we propose and integrate a novel approximate strategy into our defense, which addresses a critical computational bottleneck of defense methods and improves defense efficiency. Through extensive empirical evaluations across various realistic model-sharing scenarios, we demonstrate that our method outperforms existing defenses by more effectively defending against CPI attacks, preserving ML model utility, and reducing computational overhead.
}

\maketitle
\thispagestyle{empty} 
\setcounter{page}{0}
\DoubleSpacedXI
\section{Introduction}
\label{sec:intro}
Driven by rapid technological advancement and increasing complexity of business environments, the application of Machine Learning (ML) in business is no longer a distant aspiration but has become a strategic imperative. 
Firms across industries are increasingly turning to ML to streamline operations, personalize customer experiences, and unlock data-driven insights \citep{huang2018artificial,pant2015web,fang2019prescriptive, zhao2023exploiting, zhang2024proactive}.
However, the path to ML implementation remains challenging, particularly for firms lacking in-depth technical skills or substantial computational resources~\citep{ransbotham2017reshaping}. In reality, developing an effective ML model is far from trivial; it requires vast quantities of high-quality training data, specialized expertise, and significant computational infrastructure~\citep{priestley2023survey, thompson2023importance}. Even constructing a high-quality training dataset, often the very first step of ML model training, can typically incur costs ranging from \$10,500 to \$85,000.\footnote{See \href{https://hackernoon.com/machine-learning-costs-price-factors-and-real-world-estimates}{https://hackernoon.com/machine-learning-costs-price-factors-and-real-world-estimates}  (last accessed on Aug 18, 2025).} 
The total cost required to develop an in-house ML model can easily exceed 
one million US dollars.\footnote{See \href{https://itrexgroup.com/blog/machine-learning-costs-price-factors-and-estimates/}{https://itrexgroup.com/blog/machine-learning-costs-price-factors-and-estimates/} (last accessed on Aug 18, 2025).} 
As a result, not all firms have the capability to develop their own ML models. This challenge is particularly acute for small and medium-sized enterprises (SMEs) with limited funds and resources. Worse still, the inability of these SMEs to train ML models makes it difficult for them to compete with industry giants that are fully empowered by ML, placing SMEs at risk of bankruptcy and intensifying market monopolization~\citep{bammens2021midsize}.

To address this significant issue, a new business mode known as \textit{model-sharing} has emerged as a viable solution.  Specifically, model-sharing facilitates collaboration between two key stakeholders: the model provider, who develops high-quality ML models and shares them for profit, and the model user, who pays for the usage rights of these models. A real-world example of this model-sharing business can be found in Databricks\footnote{See \href{https://www.databricks.com}{https://www.databricks.com} (last accessed on Aug 18, 2025).}, a well-known platform for building, deploying, and sharing enterprise-grade ML solutions. Databricks allows model providers to upload and monetize their well-trained models through the Databricks Marketplace, where other firms can pay for access to these models, either by downloading the models directly or querying them via APIs, for their own use.
Through the model-sharing business, both stakeholders stand to benefit. On the one hand, model users, typically firms lacking capability to train ML models, can access advanced ML capabilities via partnering with model providers at relatively low cost, without bearing the substantial expenses associated with developing their own ML models from scratch. On the other hand, model providers, which are firms that have already trained effective ML models, can earn revenue by sharing these models, offset model training costs, and accumulate funds for developing next-generation ML models as ML technology rapidly evolves~\citep{zhao2018packaging, gan2023model}.

Although model-sharing could yield significant benefits, numerous firms remain hesitant to share their models for profit. Particularly, the concern over data confidentiality continues to pose a significant barrier, discouraging firms from monetizing their ML models~\citep{carlini2019secret, gan2023model}. Such concerns are not unfounded.
Recent studies on \textit{Confidential Property Inference} (CPI) attacks have shown that even when model providers share only their ML models but not the training data, certain confidential properties of the training data are still at risk of disclosure~\citep{ganju2018property, zhang2021leakage, mahloujifar2022property}. These confidential properties are often descriptive statistics of attribute distributions across the entire dataset, such as the loan default rate for a bank, which is the proportion of users who default on their loans. These properties are highly sensitive as they often reveal critical insights into a company's operational strategies and competitive advantages. If these properties are leaked, the model provider could lose its competitive edge and face significant operational disruptions. Using a bank's loan default rate as an example, disclosure of a high loan default rate could undermine public trust in the bank, potentially leading to a bank run and ultimately causing the bank to fail~\citep{diamond1983bank}. 

Therefore, protecting model providers from CPI attacks has become a critical prerequisite for the broader adoption of model-sharing.  Recent research efforts have been devoted to designing defense methods to address this issue. Existing defenses can be categorized as noise-based methods and model-based methods. In general, noise-based methods aim to obscure confidential properties by injecting noise at various stages of the model training process, thereby preventing adversaries from extracting sensitive attributes from the trained model through CPI attacks (e.g.,~\citealp{mahloujifar2022property}). Exemplar studies such as~\citet{ganju2018property} and~\citet{suri2023dissecting} propose adding noise to the training data, while other works explore injecting noise into model gradients~\citep{mahloujifar2022property}, intermediate embeddings, or final outputs~\citep{wang2022group}. However, the introduction of noise inevitably degrades the utility of the resulting ML models~\citep{jayaraman2019evaluating}. Such reduced model utility directly undermines the practical and commercial value of ML models, thereby rendering noise-based methods inappropriate for 
real-world model-sharing business.
To overcome this issue, recent research has shifted toward model-based methods. Typically, these defense methods first simulate a CPI adversary and allow it to attack the ML model intended for sharing~\citep{stock2022lessons, stockproperty}. Next, the ML model parameters are adjusted to minimize the attack’s effectiveness while preserving as much ML model utility as possible. Concretely, these defense methods extend the original training objective of the ML model, which typically focuses solely on maximizing ML model performance, to additionally account for minimizing the effectiveness of the simulated CPI attack~\citep{stockproperty, noorbakhsh2024inf2guard}. 
In doing so, the resulting ML model becomes robust to CPI attacks, capable of effectively protecting providers from CPI attacks while maintaining overall utility.

However, existing model-based defense methods still face two fundamental limitations in protecting model providers against CPI attacks in real-world scenarios. \textit{First}, these methods commonly assume that adversaries are non-adaptive to the specific model they are targeting.
This assumption overlooks a key characteristic of real-world adversaries: their responsiveness, that is, adversaries' ability to dynamically adjust their attack models based on the specific target and its defenses~\citep{song2021systematic}. Such responsiveness of adversaries would inevitably invalidate defense methods that assume adversaries are non-adaptive. Therefore, an effective defense method must account for the responsive nature of real-world adversaries. Yet, none of the existing model-based defenses are capable of doing so.
\textit{Second}, current model-based defense methods generally involve high computational complexity in producing robust models, primarily because the first step of the defense process, i.e., simulating CPI adversaries, incurs substantial computational cost \citep{suri2023dissecting}. This high computational complexity often makes these defense methods time-consuming to apply. 

In response to the above challenges, we propose a novel model-based defense method.
This method explicitly accounts for the responsive nature of real-world adversaries and significantly improves the computational efficiency in defense, thereby addressing two critical limitations of existing model-based defenses: the assumption of non-adaptive adversaries and the high computational overhead.
To achieve this, we first devise a Responsive CPI attack model, which allows the adversary to adapt its attack model based on the target ML model. This Responsive CPI attack model thus emulates the responsive behaviors of adversaries that shared ML models may encounter in real-world scenarios. Next, to defend against this responsive attack, we introduce an attack–defense arms race framework, where both the attack model and the defense are iteratively improved. This iterative process ultimately yields an ML model that is robust to responsive CPI attacks, thereby effectively protecting the confidentiality of the model provider’s training data during model-sharing. Moreover, we propose an approximation technique that significantly reduces the computational costs associated with attack simulation, ensuring the practicality of our defense method. Taken together, our study makes three methodological contributions: a Responsive CPI attack model for emulating real-world responsive adversaries, an arms race framework for developing effective defenses against responsive adversaries, and an approximation strategy for efficiently simulating CPI adversaries in the defense process. Extensive empirical evaluations demonstrate that our proposed defense method significantly outperforms existing benchmarks in defending against CPI attacks, but also maintains the utility of the ML model while reducing computational time.

\section{Literature Review}
\label{sec:lit_rev}
In this section, we first review three streams of research related to our study: privacy attacks on ML models, CPI attacks threatening model-sharing business, and existing defense methods against CPI attacks. We then identify the research gaps and highlight the key novelty of our study.

\subsection{Privacy Attacks on ML Models}

Our study falls into the domain of Information Privacy, a research area that has received substantial attention from Information Systems (IS) scholars \citep{smith2011information,li2011protecting, li2017anonymizing,xu2022guest}. Within this domain, one line of IS research empirically investigates individuals’ or organizations’ information disclosure behaviors, aiming to explain how they perceive, evaluate, and respond to risks related to confidential information disclosure and associated government regulations \citep{xu2012research,buckman2019relative,xu2022contextualizing}. For instance, \citet{xu2012research} examine how different privacy assurance methods affect context-specific concerns for information privacy and find that perceived control over personal information is a key determinant.
Another line of IS research focuses on developing technical methods to protect information privacy when sharing data for new business opportunities    \citep{li2011protecting,li2017anonymizing,han2025data}. For example, \citet{li2017anonymizing} propose a clustering-based anonymization approach to enable privacy-preserving sharing of medical text records in health information systems. 
Our study contributes to the latter line of work by addressing an information privacy challenge that arises when ML model owners share their models.

Privacy concerns surrounding ML models have recently received widespread attention. Indeed, privacy attacks on ML models have become increasingly prevalent and can be broadly classified into four types: model extraction attacks, membership inference attacks, reconstruction attacks, and confidential property inference (CPI) attacks \citep{rigaki2023survey}. 
Specifically, model extraction attacks seek to duplicate the ML model by recovering its parameters or architecture through query access to the target model \citep{tramer2016stealing, oh2018towards}. For example, \citet{tramer2016stealing} demonstrate that by carefully designing queries to a linear or logistic regression model and then analyzing its returned confidence scores, an adversary can accurately recover parameters of the model. In contrast, membership inference attacks aim to determine whether a specific individual’s data is included in the training set \citep{shokri2017membership, salem2018ml}. Such attacks are particularly critical in the healthcare domain, as inferring whether an individual's data appears in a private medical dataset could expose the individual's medical conditions, constituting a significant privacy violation~\citep{hu2022membership}. \citet{shokri2017membership} exemplify this attack by training a classifier on the posterior probabilities produced by the target ML model to differentiate members from non-members of the training dataset.
Unlike membership inference attacks focusing solely on confirming an individual's presence in the training set, reconstruction attacks take the threat a step further by recovering sensitive attributes of that individual \citep{fredrikson2015model, jia2018attriguard}. For instance, \citet{fredrikson2014privacy} show that given a patient's non-sensitive attributes, such as height, age, and weight, it is possible to reconstruct the patient’s sensitive genetic traits from the outputs of an ML model trained for pharmacogenetic prediction.
Lastly, CPI attacks target the extraction of confidential properties from the training datasets of ML models~\citep{ateniese2015hacking, melis2019exploiting}. These properties are typically descriptive statistics of attribute distributions across the entire dataset. For firms that train ML models using proprietary data, such properties are highly sensitive, as they potentially reveal trade secrets and strategic information~\citep{ganju2018property}. Consequently, CPI attacks constitute a significant obstacle for ML model-sharing businesses, which we review next.

\subsection{Confidential Property Inference (CPI) Attacks}
\label{subsec: attack_review}

Existing research has demonstrated that adversaries can infer confidential properties of training data even when their access is limited solely to trained ML models.  Such attacks are referred to as Confidential Property Inference (CPI) attacks. A seminal study by \citet{ateniese2015hacking} introduces the first CPI attack, operating under the assumption that adversaries have full knowledge of the target ML model's architecture and parameters, a scenario later defined as the ``white-box setting.'' To execute this attack, \citet{ateniese2015hacking} propose training numerous surrogate models (commonly known as shadow models) using datasets with varying confidential properties. The parameters of these shadow models, along with their associated confidential properties, form the training data for the white-box CPI attack model. By learning the correlation between model parameters and confidential properties, this attack model can then use the parameters of a targeted ML model to infer confidential properties present in the target model's training data.  Although this shadow-model-based CPI attack framework has been widely adopted in subsequent studies, the significant computational overhead involved in training numerous shadow models has been consistently highlighted \citep{suri2023dissecting, bertran2023scalable, chaudhari2023snap}. 

Built on \citet{ateniese2015hacking}, \citet{ganju2018property} further enhance CPI attacks specifically focusing on deep learning models, characterized by extensive parameters (e.g., neuron weights). Concretely, they propose a systematic organization of parameters whereby neurons are first sorted according to the sum of their associated weights, followed by sorting the weights within each neuron by their magnitude. Such reorganization helps the CPI attack model better learn the correlation between model parameters and confidential properties, significantly improving the attack's success rate on deep learning models.
However, rising concerns over intellectual property protection have made firms more cautious, limiting adversaries’ ability to obtain the full model architectures and parameters needed for white-box attacks.  In response, \citet{zhang2021leakage} explore CPI attacks under the more practical ``black-box setting,'' where adversaries can only query the target ML model and observe its outputs. Unlike the white-box scenario, \citet{zhang2021leakage} do not utilize model parameters directly; instead, they query shadow models with a predefined query set and combine the resultant outputs with their corresponding confidential properties to create training data for the black-box CPI attack model. This enables the CPI attack model to learn correlations between ML model outputs and confidential properties of the ML models' training data, thereby facilitating CPI attacks in the black-box setting. Extending this line of research, \citet{mahloujifar2022property} develop a black-box CPI attack for collaborative learning environments. In this scenario, adversaries act as malicious collaborators and inject poisoning data during joint ML model training. 
Besides, \citet{wang2022group} focus on inferring confidential properties in social network data, such as node group properties (e.g., whether the network contains more female nodes than male nodes) and link group properties (e.g., whether there are more links between male users compared to female users). All these studies follow a general paradigm of CPI attacks.
We provide details of this paradigm in Section~\ref{subsec:preliminary_cpi}.

\subsection{Defense Methods Against CPI Attacks}
\label{subsec:defenses review}
Given the significant risk posed by CPI attacks, recent research has increasingly emphasized developing effective defenses. Extant defense methods can be broadly classified into two categories: noise-based and model-based methods.

Noise-based defense methods obscure confidential properties by introducing noise at various stages during the ML model training. A straightforward approach is to perturb the training dataset itself. For instance, \citet{ganju2018property} flip the confidential labels of a subset of training samples, whereas \citet{Junhao2022GAN} and \citet{suri2023dissecting} modify the dataset by removing or adding samples. Another line of work injects noise into the gradients during training. Exemplar studies by  \citet{mahloujifar2022property} and  \citet{chaudhari2023snap} employ differential-privacy mechanisms to perturb gradients during training, thereby reducing an adversary’s ability to infer confidential properties from parameter updates. Additionally, \citet{wang2022group} propose injecting differentially private noise directly into model parameters in the white-box setting or into model outputs in the black-box setting.
However, noise-based defense methods inevitably degrade the performance of the ML model due to the noise injected during training. Particularly, such a reduction in model utility directly undermines the practical and commercial value of ML models, making noise-based defenses unsuitable for model-sharing business.  

Unlike noise-based defenses that indiscriminately introduce noise into the training process, model-based defense methods strategically modify ML model parameters to diminish the effectiveness of a pretrained attack model. Concretely, these defense methods typically begin by simulating a CPI adversary and training an attack model that aims to infer confidential properties either from the target model’s parameters (in the white-box setting) or from its query outputs (in the black-box setting). The target model’s parameters are then deliberately retrained to minimize the objective function of such attack model (e.g., the inference accuracy). In this vein, \citet{stock2022lessons} propose a property unlearning algorithm, which calculates gradients with respect to ML model parameters, nudging to decrease the accuracy of a well-trained CPI attack proposed by \citet{ganju2018property}. The algorithm then carefully selects a suitable learning rate to apply these gradients, ensuring a reduction in the adversary’s ability to infer confidential properties. However, while this approach can effectively diminish the success of the simulated CPI adversary, it still does not guarantee preservation of the ML model’s utility. To this end, \citet{stockproperty} further propose retraining the ML model with a combined objective that minimizes the inference accuracy of the simulated CPI attack while preserving as much ML model's utility as possible. By optimizing this joint objective, the retrained ML model can achieve robustness against that CPI attack while maintaining high predictive performance.

Our literature review suggests the following research gaps. \textit{First}, existing model-based defense methods typically assume a non-adaptive attack model, meaning that the attack model is trained independently of the target model and remains unchanged throughout the defense optimization process. In reality, however, adversaries frequently adjust their attack models dynamically in response to the specific information of the targeted ML model and its defense.  As a result, current model-based defenses overlook the responsive nature of real-world adversaries, inevitably rendering such defenses ineffective in practice. To address this gap, we propose a novel model-based defense method which explicitly accounts for the responsive nature of real-world adversaries. In doing so, we first develop a Responsive CPI attack, which effectively emulates the responsive behavior of real-world adversaries---a capability not achieved by any existing CPI attack methods. Next, we introduce an attack-defense arms race framework that iteratively modifies the ML model, thereby enhancing its robustness against responsive CPI adversaries.   \textit{Second}, recall that simulating an attack model is a necessary first step in model-based defense methods. However, this step is computationally intensive due to the requirement of training numerous shadow models, making existing model-based defenses highly time-consuming to implement. Such computational overhead significantly limits the practical applicability of current defenses in real-world ML model-sharing scenarios.
To this end, we propose an approximation technique that substantially reduces the computational costs associated with attack model simulation, ensuring the practicality of our defense method. In summary, the proposed Responsive CPI attack, the arms race-based defense framework, and the approximation strategy for computational efficiency together constitute the key methodological contributions of this study.

\section{Problem Formulation and Preliminaries}
\label{sec:prob_for}

We formally define the secure model-sharing problem in Section \ref{subsec:problem}, and present preliminaries on the general paradigm of existing CPI attacks in Section~\ref{subsec:preliminary_cpi}, which serves as the technical foundation of our study.

\subsection{Problem Formulation}
\label{subsec:problem}

Consider a model provider intending to share a well-trained ML model $f_\theta: \mathcal{X}\xrightarrow{}\mathcal{Y}$, parameterized by $\theta$ and trained on the provider's proprietary dataset $D=(X,Y)$. Here, $X$ is a matrix in which each row represents the attribute values of an instance, with each instance belonging to the input space $\mathcal{X}$. Correspondingly, $Y$ is a vector of labels for these instances, associated with the target prediction task, where each label in $Y$ belongs to the output space $\mathcal{Y}$.
For instance, a bank may wish to share its financial product telemarketing model $f_\theta$. In this case,  $X$ may include demographic characteristics and credit histories of its clients, and $Y$ indicates whether a client purchases financial products or not.
Notably, even though the model provider shares only the trained ML model (or query access to the model) without disclosing the proprietary dataset $D$, confidential properties of the dataset (as defined in Definition~\ref{def:conf_p}) can still be disclosed through Confidential Property Inference (CPI) attacks (Definition~\ref{def:cpi}).

\begin{definition}(\textbf{Confidential Property})
\label{def:conf_p}
A confidential property of the training dataset $D$ is defined as the result of applying an aggregate function to the vector $X_p$ of a specific confidential property-related attribute $p$, which corresponds to one column of attribute matrix $X$. Formally, let $P$ denote a confidential property and $\text{AGG}(\cdot)$ denote an aggregate function (e.g. mean). 
Then, $P:=\text{AGG}(X_p)$.\footnote{In this study, we focus exclusively on categorical attributes, which is a common setting in existing works \citep{ganju2018property, zhang2021leakage}.}
\end{definition}

For example, consider a bank's dataset containing client attributes such as demographic characteristics (e.g., age and job), along with financial information indicating whether clients hold loans and their corresponding default statuses. The attribute matrix can be thus represented as $X=[ X_\textit{age}, X_\textit{job}, X_\textit{loan}, X_\textit{default}]$. Since the default rate reflects the bank's effectiveness in risk management and overall asset quality, competitors gaining access to this information could strategically undermine the bank’s market position and competitive advantage~\citep{agoraki2011regulations}. Hence, the default rate of clients is one confidential property of this dataset, i.e., $P= \text{MEAN}(X_\textit{default})$,
with the corresponding confidential property-related attribute being $\textit{default}$.

\begin{definition} (\textbf{Confidential Property Inference (CPI) Attack})
\label{def:cpi}
Consider an ML model $f_\theta: \mathcal{X}\xrightarrow{}\mathcal{Y}$ as the target model, parameterized by $\theta$ and trained on a dataset $D=(X,Y)$. A Confidential Property Inference (CPI) attack aims to infer a confidential property $P$ of training data $D$, using information accessible from the target model. This information may consist of the target ML model's parameters in the white-box setting or the model’s outputs on a predefined query set in the black-box setting. Precisely,  let $\mathcal{F}_\theta$ denote the accessible information from the target model $f_\theta$. A CPI attack trains an attack model $h_\phi$ parameterized by $\phi$, with the objective of estimating the confidential property $P$ from $\mathcal{F}_\theta$, i.e.,  $\hat{P} = h_\phi(\mathcal{F}_\theta)$.
\end{definition}

As a result, given the threat posed by CPI attacks, when a model provider intends to share its ML model for profit, it becomes imperative to adjust the model such that it is CPI-attack robust, that is, the confidential properties of the models' training data cannot be inferred through CPI attacks. Moreover, it is noted that a shared model holds commercial value only if it maintains high utility (i.e., target predictive performance). Therefore, the adjusted ML model, while being robust to CPI attacks, should also preserve as much utility as possible.  Now, we are ready to formally define our secure model-sharing problem. 

\begin{definition} \label{def:sms} (\textbf{Secure Model-Sharing (SMS) Problem})
A model provider intends to share its ML model $f_\theta$ for profit, where $f_\theta$ is trained on the model provider's proprietary dataset $D$ that contains a confidential property $P$. Given the threat posed by CPI attacks, the model provider needs to adjust the model's parameters $\theta$ prior to sharing, with the following two objectives: 
\begin{itemize}
    \item{\textbf{CPI-attack robustness.}} The confidential property $P$ of the training data cannot be inferred by adversaries through CPI attacks when they have access to the model information $\mathcal{F}_\theta$ after the model is shared.
    \item{\textbf{Utility retention.}} The shared ML model should retain as much utility as possible for the target prediction task.
\end{itemize}
\end{definition}

\subsection{Preliminary: General Paradigm of CPI Attacks}
\label{subsec:preliminary_cpi}
In this section, we introduce the general paradigm of existing CPI attacks. 
The realization of CPI attack is grounded in the fact that ML models trained on analogous datasets tend to exhibit similar characteristics, such as similar parameters or outputs~\citep{ateniese2015hacking, ganju2018property}. 
Such fact implies the existence of a learnable mapping from a model’s information $\mathcal{F}_\theta$ (e.g., model parameters) to the confidential properties of its training dataset. CPI attacks aim to learn this mapping by training an attack model $h_\phi$.

\begin{figure}[t]
\centering
\includegraphics[width=16cm]{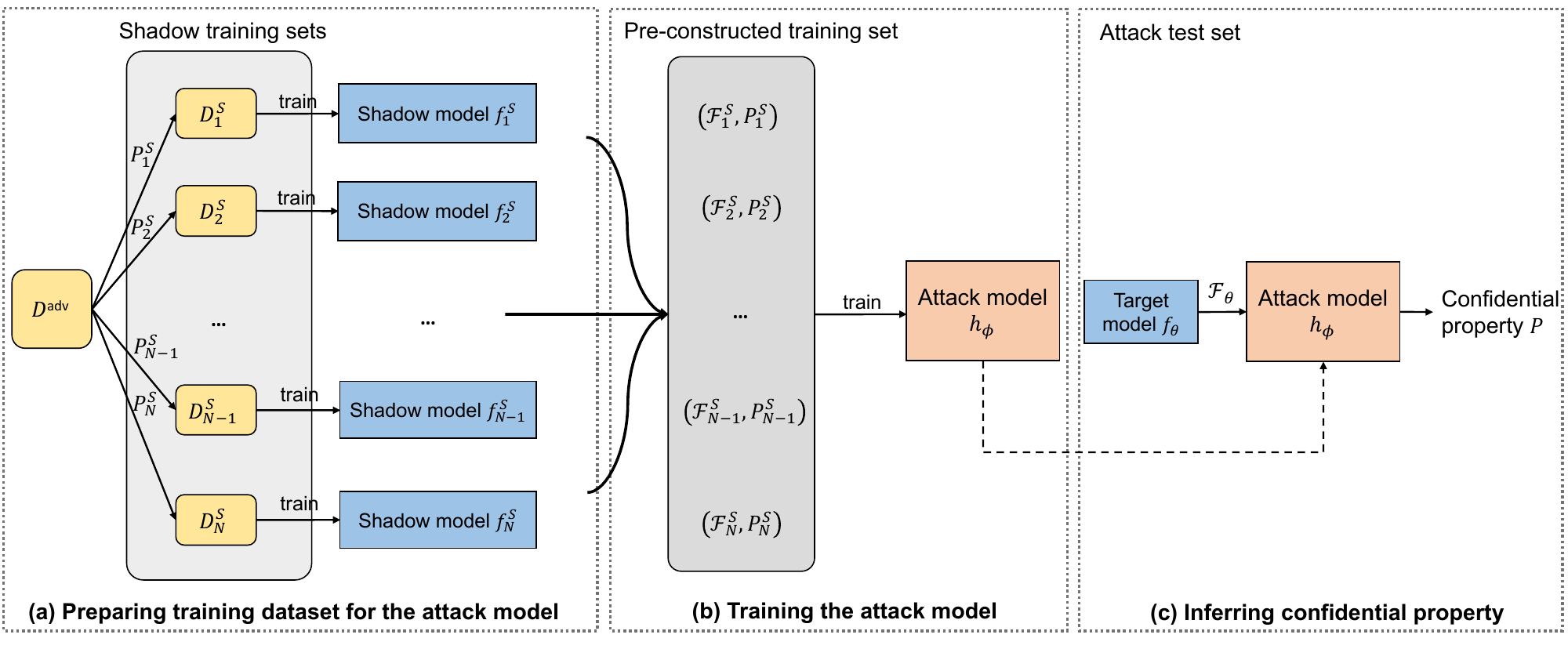}
\vspace{-10pt}
\caption{A common CPI attack paradigm consists of three phases: (a) preparing the training dataset for the attack model, (b) training the attack model, and (c) inferring confidential properties from the target model's training dataset.}
\vspace{-5mm}
\label{fig:attack paradigm}
\end{figure}

To facilitate this, an appropriate training dataset must be constructed, consisting of training samples that include model parameters or outputs along with their corresponding confidential properties. Once the attack model is well-trained, it can be utilized to infer the confidential property in the target model's training dataset. Therefore, the commonly used paradigm for conducting such a CPI attack consists of three phases: (1) preparing the training dataset for the attack model $h_\phi$, (2) training the attack model $h_\phi$, and (3) inferring the confidential property using the trained $h_\phi$. The pipeline of a CPI attack is illustrated in Figure \ref{fig:attack paradigm}.

\textit{PHASE 1: Preparing the Training Dataset for the Attack Model $h_\phi$}. To effectively train the attack model, the adversary must first prepare the attack model's training dataset, where the $i$-th training sample $(\mathcal{F}^S_i, P^S_i)$ includes model information $\mathcal{F}^S_i$ (i.e., the model parameters or outputs on a predefined query set) of a trained ML model, along with the corresponding confidential property $P^S_i$ of that ML model's training data. 
Existing studies have shown that an adversary can obtain an auxiliary dataset $D^{\text{adv}}=(X^{\text{adv}},Y^{\text{adv}})$ that shares the same attribute space and the target label space as the target ML model's training data \citep{shokri2017membership, nasr2018comprehensive, song2021systematic}.\footnote{
This is because a model user would only purchase and use the provider’s ML model if they have input data with attributes and target label compatible with the model. Therefore, the attribute space of the provider’s training data is known to the model user and can be easily obtained by the adversary. The actual attribute values for $D^{\text{adv}}$ can be further gathered from publicly available sources or the adversary’s private data \citep{ganju2018property, zhang2021leakage,Junhao2022GAN, mahloujifar2022property}. } 
This auxiliary dataset $D^{\text{adv}}$ allows the adversary to systematically construct the training data for the attack model.

Specifically, to construct a single training sample  $(\mathcal{F}^S_i, P^S_i)$, the adversary first randomly samples a subset of data $D_i^S\subset D^{\text{adv}}$. 
This sampled dataset $D_i^S$ is then used to train an ML model $f_i^S$, commonly referred to as a \textit{shadow model}~\citep{shokri2017membership, ganju2018property, zhang2021leakage}.\footnote{Existing literature on CPI attacks commonly assumes that the adversary has knowledge of the target model’s architecture~\citep{ateniese2015hacking, ganju2018property, wang2022group}. Accordingly, the shadow model $f_i^S$ is constructed using the same architecture as the target model. In white-box settings, this assumption is straightforward, as the adversary has direct access to the model parameters and architecture. Even in black-box settings, the model architecture can often be inferred through model architecture extraction techniques prior to conducting CPI attacks \citep{oh2018towards, wang2018stealing}.} Subsequently, the adversary derives the shadow model's information $\mathcal{F}^S_i$ and computes the confidential property $P^S_i$ of the shadow model's training dataset $D_i^S$.\footnote{In the white-box setting, the accessible model information is model parameters. When the parameters are not naturally in vector form (e.g., in deep neural networks where weights span multiple layers), transformations proposed by \citet{ganju2018property} are applied to convert them into a vector format. } The resulting pair $(\mathcal{F}^S_i, P^S_i)$ constitutes one training sample for the attack model $h_\phi$. By repeating this procedure for $N$ shadow models, the adversary ultimately compiles the training dataset for the attack model, represented as $\{(\mathcal{F}^S_i, P^S_i) \}_{i=1}^N$.

\textit{PHASE 2: Training the Attack Model $h_\phi$.}
Given the training dataset $\{(\mathcal{F}_i^S, P^S_i) \}_{i=1}^N$, the adversary aims to train the attack model $h_\phi$ with learnable parameters $\phi$ by optimizing
\begin{equation*}
\min_\phi \sum_{i=1}^N l(h_\phi(\mathcal{F}^S_i), P^S_i),
\end{equation*}
where $l(\cdot,\cdot)$ is a loss function that quantifies the error of the attack model $h_\phi$ in predicting the confidential property $P^S_i$ from the corresponding model information $\mathcal{F}^S_i$.

\textit{PHASE 3: Inferring Confidential Property.}  
With the well-trained attack model $h_\phi$, the adversary can now infer the confidential property of the target ML model’s training data, treating the target model’s information as a test sample for the attack model. Precisely, given the target ML model's information $\mathcal{F}_\theta$, the predicted confidential property $\hat{P}$ is computed as
\begin{equation*}
\hat{P}=h_\phi(\mathcal{F}_\theta).
\end{equation*}

\section{Solution Method: $\text{D}$-$\text{S}^2$HARE}
\label{sec:method}
In this section, we propose a novel defense method to address the SMS problem. Recall that the model provider must account for two key objectives when adjusting the ML model parameters: (1) ensuring CPI-attack robustness, and (2) retaining the ML model’s utility for its target prediction task. Notably, our literature review has shown that model-based defense methods provide a viable approach to simultaneously achieving both objectives \citep{stock2022lessons}. Our proposed method thus follows the model-based defense paradigm.

More precisely, model-based defense methods begin by simulating a CPI adversary and training an attack model $h_\phi$ following the paradigm introduced in Section~\ref{subsec:preliminary_cpi}, which is capable of inferring the confidential property $P$ from the ML model’s information $\mathcal{F}_\theta$. Once such an attack model is trained, the defense proceeds by adjusting the target ML model’s parameters $\theta$ 
to maximize a composite objective function that combines the error of the attack model in predicting the confidential property, denoted by $\mathcal{L}_\mathcal{P}$, and the utility of the ML model, denoted by $-\mathcal{L}_\mathcal{T}$. 
The resulting 
model parameters are thus obtained by solving the following optimization problem:
\begin{equation}
\label{eq:obj_def}
    \arg\max_{\theta} \left[ \mathcal{L}_\mathcal{P}(P, h_\phi(\mathcal{F}_\theta)) -\lambda \mathcal{L}_\mathcal{T}(\theta;D) \right],
\end{equation}
where hyperparameter $\lambda$ controls the relative importance of these two loss terms.

\begin{figure}[t]
    \centering
    \includegraphics[width=17cm]{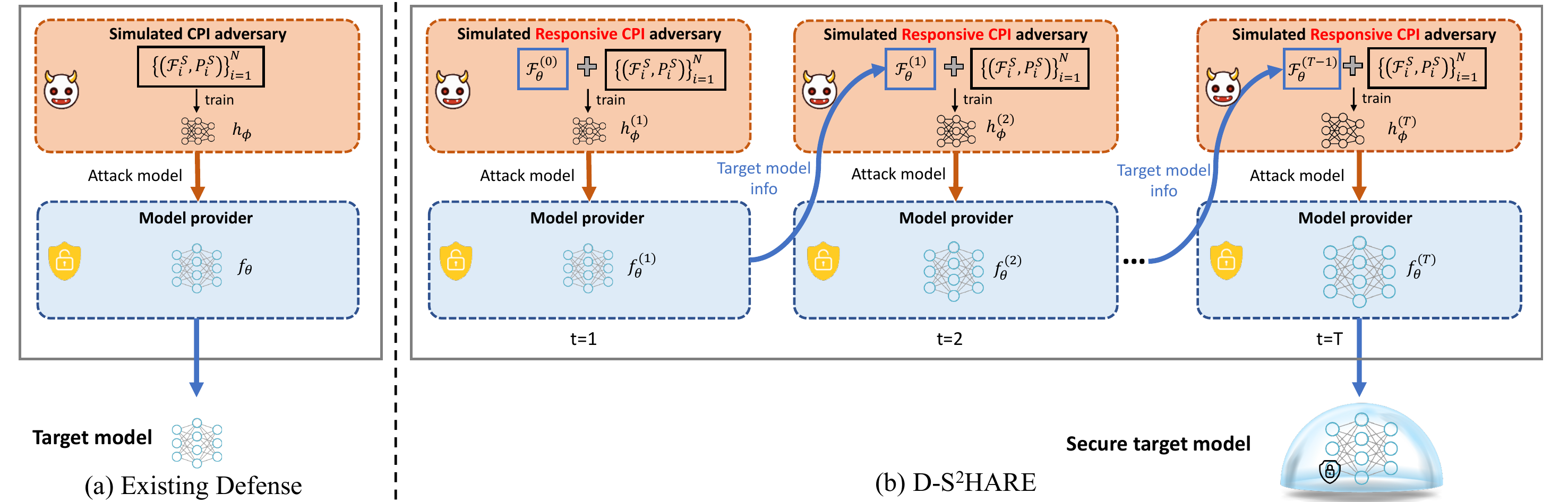}
    \caption{Comparison of Existing Defenses with Our Proposed $\text{D}$-$\text{S}^2$HARE}
    \label{fig:framework}
\end{figure}

However, because these existing model-based defenses simulate the attack model in advance, they implicitly assume that the attack is static and trained independently of the target model. This neglects a critical characteristic of real-world adversaries: their responsiveness. In practice, adversaries often adapt their attack model based on the information of the target ML model $\mathcal{F}_{\theta}$, in order to improve attack effectiveness. Overlooking such responsiveness can thus result in ineffective defenses. 
To address this limitation, we first develop a Responsive CPI attack that effectively reflects the responsive behavior of real-world adversaries. Building on this, we propose an attack-defense arms race framework that iteratively refines the ML model to be robust against responsive adversaries. Figure~\ref{fig:framework}
illustrates this key difference between our proposed defense method and existing model-based defenses.
Additionally, considering the significant computational overhead in simulating attack models, which is primarily due to the need to train numerous shadow models, we propose an approximation technique to reduce the computational cost associated with simulating the attack model, thus enhancing the practicality and efficiency of our defense method in realistic model-sharing scenarios. 

These three innovations, the Responsive CPI attack, the arms race-based defense framework, and the approximation technique, constitute our novel model-based defense method for addressing the SMS problem. Together, they enable our proposed defense method to effectively counter responsive adversaries while ensuring high defense efficiency.
We detail them in Sections~\ref{subsec: responsive_attack}, \ref{subsec: arms race}, and~\ref{subsec:approximation}, respectively. Accordingly, we designate our defense method as $\text{D}$-$\text{S}^2$HARE, which stands for a \underline{D}efense method for the \underline{S}ecure model-\underline{SH}aring problem that considers \underline{A}dversarial \underline{R}esponsiveness and defense \underline{E}fficiency. The complete defense procedure for  $\text{D}$-$\text{S}^2$HARE is summarized in Section~\ref{subsec:overall alg}.

\subsection{Responsive CPI Attack}
\label{subsec: responsive_attack}
Recall that real-world adversaries are highly responsive, constantly adapting their attack models to changes in the target models.
Therefore, our proposed CPI attack must be capable of capturing this responsiveness.
Moreover, since adversaries seek to effectively extract confidential properties, our simulated attack also requires prioritizing inference effectiveness. These two fundamental requirements: responsiveness and effective inference, establish the objectives of our attack design, ensuring that it precisely reflects real-world adversaries. 
In what follows, we begin by discussing why current CPI attacks fail to satisfy these two requirements, then introduce our proposed Responsive CPI attack, which is grounded in a rigorous theoretical foundation.

\vpara{Limitations of existing CPI attacks.}
As discussed in Section~\ref{subsec: attack_review}, existing CPI attacks train attack models by minimizing the empirical error on training data, with the expectation that these attacks will effectively infer the confidential properties of unseen target ML models.
More precisely, an attack model is trained on a pre-constructed dataset $\{(\mathcal{F}^S_i, P^S_i) \}_{i=1}^N$, consisting of $N$ training samples. 
For a given training instance $(\mathcal{F}^S_i, P^S_i)$, $\mathcal{F}^S_i$ represents the $i$-th shadow model $f_i^S$'s model information (i.e., parameters or outputs), which follows the distribution $\mathcal{D}_S$. $P_i^S$ represents the confidential property associated with the training dataset of shadow model $f_i^S$.
The training objective of the attack model $h_\phi$ is to minimize the empirical error $\widehat{R}(h_\phi)$, which is given by:
\begin{equation*}
\widehat{R}(h_\phi)=\frac{1}{N} \sum_{i=1}^N l(h_\phi(\mathcal{F}^S_i),P^S_i),
\end{equation*}
where $l(h_\phi(\mathcal{F}^S_i),P^S_i)$
denotes the loss measuring the discrepancy between the predicted confidential property  $h_\phi(\mathcal{F}^S_i)$ and the true confidential property $P^S_i$ in the $i$-th training sample.

Once trained, the attack model is expected to effectively infer the confidential property of a target ML model, given the target model's information $\mathcal{F}_\theta$.  The target model thus serves as a test instance for the well-trained attack model $h_\phi$. Accordingly, the attack model's inference effectiveness corresponds to its test error $R(h_\phi)$, computed as:
\begin{equation} \label{eq:test_error}
R(h_\phi)=E_{\mathcal{D}_T}[l(h_\phi(\mathcal{F}_\theta),P)]. 
\end{equation}
The test error $R(h_\phi)$ quantifies the expected performance of $h_\phi$ in inferring a target model’s confidential property, with the expectation taken over the distribution $\mathcal{D}_T$ of the target model’s information $\mathcal{F}_\theta$.

The above formulation illustrates that existing CPI attacks fundamentally cannot meet either of the two critical requirements: responsiveness and effective inference. First, with respect to responsiveness, the attack model in existing CPI attacks is trained to minimize the empirical error $\widehat{R}(h_\phi)$ on a pre-constructed dataset,
without incorporating any information from the target ML model. This renders the attack model's training process entirely independent of the target, resulting in an attack model that remains static and unable to adapt to the specific information of the target ML model. 
Second, existing CPI attacks fail to achieve effective inference due to their neglect of distribution shift\footnote{Here, distribution shift refers to a discrepancy between the distribution of the ML model's training data and that of its test data.} between the pre-constructed training dataset and the target ML model (i.e., the test instance). Concretely, when the distribution of model information in the pre-constructed training datasets (i.e., $\mathcal{D}_S$) differs significantly from that of the target ML model (i.e., $\mathcal{D}_T$),  the attack effectiveness inevitably deteriorates, as the attack model struggles to generalize across different data distributions. For instance, consider an adversary from one financial institution (Bank A) who trains an attack model using customer data reflecting Bank A’s specific demographic and behavioral patterns. When this adversary attempts to apply the trained model to infer the confidential property of a target ML model shared by a different financial institution (Bank B), discrepancies between the customer populations would cause substantial distribution shifts. Consequently, the attack model, trained on Bank A’s data, fails to generalize effectively to Bank B’s target model, resulting in a significant increase in test error $R(h_\phi)$  when inferring Bank B's confidential property.

To this end, we propose a novel Responsive CPI attack that  explicitly incorporates information of the target ML model into the attack model’s training process, thereby capturing the responsive nature of real-world adversaries. Meanwhile, it mitigates the distribution shift between the pre-constructed training dataset and the target model, which enhances the inference effectiveness of the attack.

\vpara{Theoretical foundation of Responsive CPI attack.} Before we delve into the detailed method design of Responsive CPI attack, we build on the work of \citet{shimodaira2000improving} to establish the following theorem, which provides the theoretical guideline for designing our attack model's training objective.
\begin{theorem} \label{theorem: weighted_loss}
    Let $h_\phi$ denote an attack model that maps an ML model's information $\mathcal{F}^S_i$ to its training data's confidential property $P_i^S$. Define $\mathcal{D}_S$ as the distribution of model information from the attack model’s pre-constructed training data, and $\mathcal{D}_T$ as the corresponding distribution for the test data (i.e., the target model's information in our problem).
    Suppose that the support of the distribution $\mathcal{D}_T$ is contained in the support of distribution $\mathcal{D}_S$. Then, the test error of the attack model can be expressed as:
    \begin{equation}
    \label{eq: distribution shift}
     R(h_\phi)=E_{\mathcal{D}_S}[\frac{1}{N} \sum_{i=1}^N r_i l(h_\phi(\mathcal{F}_i^S),P_i^S)],
    \end{equation}
    where $r_i=\frac{\mathrm{Pr}_{\mathcal{D}_T}(\mathcal{F}_i^S)}{\mathrm{Pr}_{\mathcal{D}_S}(\mathcal{F}_i^S)}$ is the training sample weight, $\mathrm{Pr}_{\mathcal{D}_T}(\mathcal{F}_i^S)$ and $\mathrm{Pr}_{\mathcal{D}_S}(\mathcal{F}_i^S)$ are the probability densities of $\mathcal{F}_i^S$ under the distributions $\mathcal{D}_S$ and $\mathcal{D}_T$.
\end{theorem}

Proof: See Appendix \ref{appendix: proof_weight}.

The theorem demonstrates that the test error of the attack model, $R(h_\phi)$, can be represented as a weighted sum of the losses over the training samples.  In particular, the training sample weight $r_i$ adjusts the contribution of each training sample based on the probability density of $\mathcal{F}_i^S$ under the training sample distribution $\mathcal{D}_S$ relative to that under the test distribution $\mathcal{D}_T$. This theoretical insight suggests that to minimize the test error under distribution shift, the training objective for the attack model should be formulated as minimizing the following weighted empirical loss  $\mathcal{L}_\mathcal{A}$:
\begin{equation}
\label{eq:La}
    \mathcal{L}_{\mathcal{A}}:= \frac{1}{N} \sum_{i=1}^N r_i \times l(h_\phi(\mathcal{F}^S_i), P^S_i),
\end{equation}
where $\mathcal{L}_{\mathcal{A}}$ corresponds to the integrand of the expected test error in Eq.~\eqref{eq: distribution shift}.

By adopting this training objective, Responsive CPI attack achieves dual capabilities: responsiveness to the target model and effective inference of confidential properties.
\begin{itemize}
    \item \emph{Responsiveness}:  The weights of the training samples in training objective \eqref{eq:La} can be adjusted based on the distribution of the target model’s information.
    Specifically, as the target model changes, the training sample weights are reassigned to better match the new target model. This reweighting ensures that the attack model remains responsive to the specific information of the target model.

    \item \emph{Effective inference}: The weighted training loss $\mathcal{L}_{\mathcal{A}}$ (i.e., Eq.~\eqref{eq:La}) is precisely the integrand of the expected test error defined in Eq.~\eqref{eq:test_error}. Therefore, as guaranteed by Theorem~\ref{theorem: weighted_loss}, minimizing $\mathcal{L}_{\mathcal{A}}$ directly contributes to reducing the test error in the presence of distribution shift, thereby improving the effectiveness of inferring confidential properties. 

\end{itemize}

\vpara{Training sample weight estimation.}  Although the key to ensuring the responsiveness of the proposed attack model is adjusting the weights $r_i$ based on the distribution of the target model’s information, calculating $r_i$ remains a significant challenge for the adversary, since the distribution of the target model’s information is typically unknown. Inspired by~\citet{sugiyama2007direct},  we derive the following theorem to estimate $r_i$ for each training sample $\mathcal{F}_i^S$ in the context of the SMS problem.
\begin{theorem} \label{theorem: r_i}
Let $K_\sigma (a, b)=exp\{-\frac{\|a-b\|^2}{2\sigma^2}\}$ denote the Gaussian kernel with kernel width $\sigma$. For the SMS problem, the weight $r_i = \frac{\text{Pr}_{\mathcal{D}_T}(\mathcal{F}_i^S)}{\text{Pr}_{\mathcal{D}_S}(\mathcal{F}_i^S)}$ for the training sample $\mathcal{F}_i^S$ can be estimated by:
\begin{equation*}
    \hat{r}_i=\frac{N\cdot K_\sigma (\mathcal{F}_i^S, \mathcal{F}_\theta)}{\sum_{j=1}^N K_\sigma (\mathcal{F}_j^S, \mathcal{F}_\theta)},
\end{equation*}
where $\mathcal{F}_\theta$ denotes the accessible model information of the target ML model.
\end{theorem}
Proof: See Appendix \ref{appendix: proof_r_i}.

With the estimated training sample weight $\hat{r}_i$, the loss for Responsive CPI attack model is formulated as:
\begin{align}
\begin{aligned}
\label{eq:La final}
    \mathcal{L}_{\mathcal{A}}&=\frac{1}{N}\sum_{i=1}^N\hat{r}_i\times l(h_\phi(\mathcal{F}^S_i), P^S_i)\\
    &= \sum_{i=1}^N \frac{K_\sigma (\mathcal{F}_i^S, \mathcal{F}_\theta)}{\sum_{j=1}^N K_\sigma (\mathcal{F}_j^S, \mathcal{F}_\theta)} \times l(h_\phi(\mathcal{F}^S_i), P^S_i).
\end{aligned}
\end{align}
Notably, this training objective explicitly incorporates the target ML model's information $\mathcal{F}_\theta$, enabling the attack model to update its parameters based on $\mathcal{F}_\theta$ and capture the responsive nature of the real-world adversary. 
Moreover, as established in Theorem~\ref{theorem: weighted_loss}, minimizing $\mathcal{L}_{\mathcal{A}}$ reduces the expected test error under distribution shift, enhancing the attack model’s effectiveness in inferring target ML model’s confidential property.

\subsection{Attack-Defense Arms Race}
\label{subsec: arms race}
Having introduced the Responsive CPI attack, which more accurately emulates real-world adversaries, we now propose a defense method against such attacks, thereby enhancing the protection of the target model’s confidential properties in practical model-sharing scenarios. 

However, defending against the Responsive CPI attack is inherently challenging. Unlike existing model-based defense methods that assume a static attack model and adjust the target ML model’s parameters accordingly, the responsive nature of real-world adversaries renders such one-shot defense strategies ineffective. Specifically, when the model provider adjusts the target ML model to defend against a particular attack model, a responsive adversary may adapt its attack in response, thereby invalidating the initial defense. As a result, static defenses fail to keep pace with adversaries that continuously adapt their attack strategies. 

To this end, we propose an arms race-based framework that mimics such iterative competition between the model provider and the adversary. This ongoing arms race enables both parties to improve continuously: the model provider iteratively refines the target model $f_\theta$ to enhance its robustness, while a simulated adversary updates its attack model $h_\phi$ according to the information $\mathcal{F}_\theta$ of the updated target model. Upon convergence of this iteration, the two sides reach an equilibrium, resulting in a target model that is robust against responsive CPI attacks.

More precisely, in the $t$-th round of iteration, given the current attack model $h_{\phi^{(t)}}$,
the model provider updates the target model's learnable parameters $\theta$ by solving the following optimization problem:
\begin{equation}
\label{eq:Lt}
    \theta^{(t)} =\arg\max_{\theta}  \left[\mathcal{L}_{\mathcal{P}}(P, h_{\phi^{(t)}}(\mathcal{F}_\theta)) - \lambda\mathcal{L}_{\mathcal{T}}(\theta;D)\right],
\end{equation}
where $\mathcal{L}_{\mathcal{P}}$ represents the test error of the attack model $h_{\phi^{(t)}}$ in inferring the confidential property $P$ of the target model given its  model information $\mathcal{F}_\theta$. Maximizing $\mathcal{L}_{\mathcal{P}}$ reduces the attack’s effectiveness, thereby strengthening the target model’s protection of its confidential property.
Moreover, $\mathcal{L}_{\mathcal{T}}$ denotes the standard training loss of the target model on its training dataset $D=(X,Y)$.\footnote{For example, let the $i$-th instance in the training dataset $D = (X, Y)$ be denoted as $(\mathbf{x}_i, y_i)$. Assuming that $y_i$ is a numerical label, the training loss $\mathcal{L}_{\mathcal{T}}$ can be defined as $\mathcal{L}_{\mathcal{T}} = \sum_i (f_\theta(\mathbf{x}_i) - y_i)^2$.}
Maximizing the negative of $\mathcal{L}_{\mathcal{T}}$ ensures that the target model retains high utility for its target task. The hyperparameter $\lambda$ controls the trade-off between CPI-attack robustness and maintaining the utility of the target model. 

Once the target model has been updated, the simulated adversary subsequently refines its Responsive CPI attack model (as illustrated in Section~\ref{subsec: responsive_attack}) based on the information of the adjusted target model,  initiating the next iteration.  Specifically, the adversary updates the learnable parameters $\phi$ of the attack model by solving the following optimization problem: 
\begin{equation*}
     \phi^{(t+1)} = \arg\min_\phi\mathcal{L}_{\mathcal{A}}(\phi, \mathcal{F}_{\theta^{(t)}}),
\end{equation*}
where $\mathcal{L}_{\mathcal{A}}$ is the training loss of the attack model, as defined in Eq.~\eqref{eq:La final}. Next, we show the convergence of this iteration of attack-defense arms race. 
\begin{theorem}
\label{theorem:convergence}
The proposed attack-defense arms race framework converges.
\end{theorem}

Proof: See Appendix \ref{appendix:proof_convergence}.

Taken together, this iterative arms race drives continuous improvements in both the model provider’s defenses and the simulated adversary’s attacks. Once convergence is reached, as guaranteed by Theorem~\ref{theorem:convergence}, the resulting secure target model $f_{\theta^{*}}$ is robust against real-world responsive adversaries while preserving its model utility, making it suitable for secure model-sharing.

\subsection{Approximation Strategy}
\label{subsec:approximation}
An essential step in the proposed defense method is simulating a responsive adversary, which must first construct a training dataset $\{(\mathcal{F}^S_i, P^S_i) \}_{i=1}^N$ to train its attack model. Here, $P^S_i$ represents the confidential property of the dataset $D^S_i$, and $\mathcal{F}^S_i$ denotes the model information of the shadow model $f^S_i$, which is trained on $D^S_i$. This dataset construction process implies that in order to obtain $N$ training samples, the adversary must train $N$ shadow models. As noted in prior work~\citep{ganju2018property, mahloujifar2022property, stockproperty}, $N$ can often reach hundreds or even thousands. Consequently, training such a large number of shadow models has emerged as a significant computational bottleneck in simulating the adversary, thereby affecting the overall efficiency of the defense method. To this end, we propose a novel approximation strategy that substantially reduces the computational cost associated with shadow model training.

\begin{figure}[!h]
	\centering
    \vspace{-5pt}
	\subfloat[Existing CPI Attacks]{
		\begin{minipage}[t]{0.4\linewidth}
			\centering
			\includegraphics[width=1.8in]{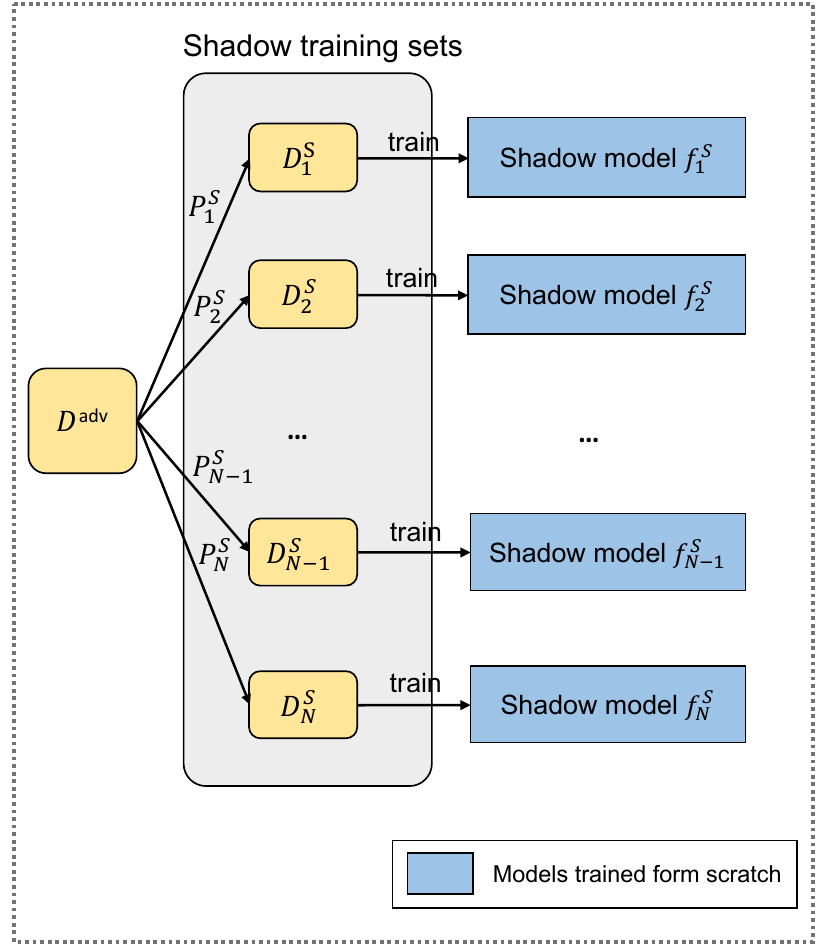}
		\end{minipage}
	}%
	\subfloat[Our Approximation Strategy]{
		\begin{minipage}[t]{0.6\linewidth}
			\centering
			\includegraphics[width=2.9 in]{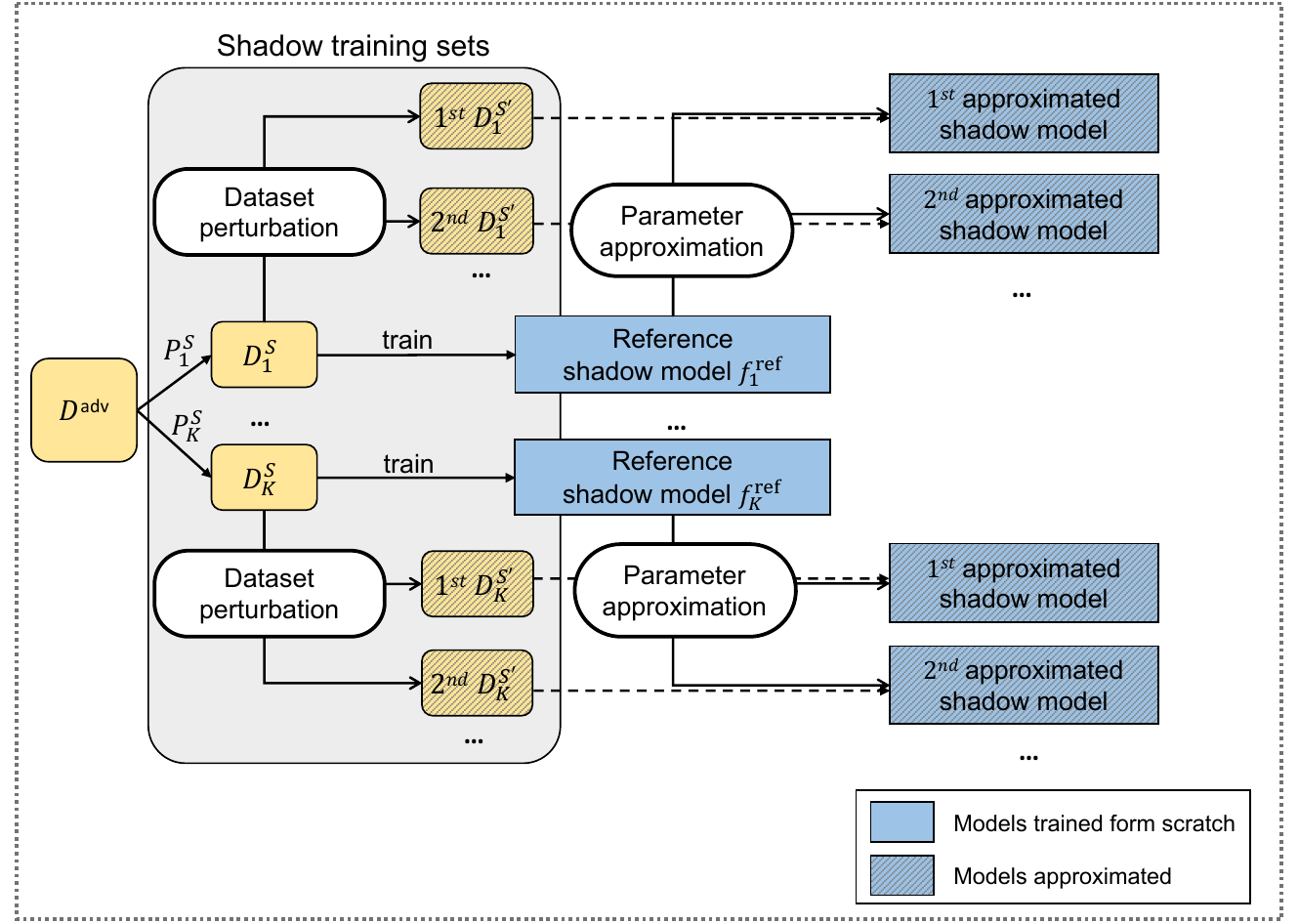}
		\end{minipage}
	}%
	\centering
    \caption{Comparison of Shadow Model Training: Existing CPI Attacks vs. Our Approximation Strategy 
    }
    \vspace{-0.5cm}
	\label{fig:approximate}
\end{figure}

Figure~\ref{fig:approximate} compares our approximation strategy with the vanilla method for training shadow models.
As illustrated, the vanilla approach samples $N$ distinct datasets and independently trains $N$ corresponding shadow models. In contrast, our proposed approximation strategy samples only $K$ datasets and trains $K$ shadow models using these datasets, where $K\ll N$. These $K$ shadow models are referred to as \emph{reference shadow models}, denoted by  $f_1^{\text{ref}}, f_2^{\text{ref}}, \ldots, f_K^{\text{ref}}$.
For the remaining $N-K$ models, instead of sampling new datasets and training each shadow model from scratch, we perturb the original $K$ datasets and estimate how the parameters of the corresponding reference shadow models would change due to these training data perturbations. This enables us to approximate the parameters of the shadow models that would have otherwise been trained on the perturbed datasets. We refer to these models, whose parameters are not obtained through actual retraining but are instead estimated based on perturbations of the training data, as \emph{approximated shadow models}.
As summarized in the overview, our approximation strategy comprises three key steps: sampling datasets and training reference shadow models, perturbing the datasets, and approximating model parameter changes due to training data perturbations. We describe each of these steps in detail below.

\vpara{Data sampling and reference shadow model training.}
Instead of training all $N$ shadow models from scratch, we train only $K$ reference shadow models $f_1^{\text{ref}}, f_2^{\text{ref}}, \dots, f_K^{\text{ref}}$. For the $k$-th reference shadow model, we begin by randomly sampling a subset of data $D_k^S=(X^S, Y^S)$ (where $X^S$ and $Y^S$ represent the attributes and the labels for the target task respectively) from the adversary's auxiliary dataset $D^{\text{adv}}$. The sampled dataset $D_k^S$ is then used to train the reference shadow model $f_k^{\text{ref}}$, which mimics the target model by learning to map attribute values $X^S$ to the corresponding labels $Y^S$ for the target task. Specifically, given the training dataset $D_k^S$ of the $k$-th reference shadow model $f_k^{\text{ref}}$, its parameters $\theta_k$ are obtained by solving the following optimization problem:
\begin{equation*}
    \theta_k = \arg\min_\theta L_{\mathcal{T}}(\theta;D_k^S),
\end{equation*}
where $L_{\mathcal{T}}(\theta;D_k^S)$ denotes the training loss of $f_k^{\text{ref}}$ evaluated on $D_k^S$.

\vpara{Data perturbation.}
Next, we perturb the datasets $D_1^S, D_2^S, \dots, D_K^S$, each of which serves as the training dataset for the corresponding reference shadow model $f_1^{\text{ref}}, f_2^{\text{ref}}, \dots, f_K^{\text{ref}}$. Specifically, given an original dataset $D_k^S$, a perturbation operation $g(\cdot)$ is applied to generate a perturbed dataset $D_k^{S^\prime}$:
\begin{equation}
\label{eq:d_per}
    D_k^{S^\prime}=g(D_k^S), g\sim G,
\end{equation}
where $G$ denotes the set of candidate perturbation operations. Each operation $g(\cdot)$ is sampled uniformly from $G$ and maps $D_k^S$ to a perturbed dataset $D_k^{S^\prime}$. In particular, we consider the following perturbation operations:

\begin{itemize}
    \item 
    \emph{Shuffling values of confidential property-related attribute}:
    This operation randomly shuffles a subset of values within the confidential property-related attribute while keeping all other attributes unchanged. Thus, it introduces greater variability into the attribute joint distribution while still preserving the confidential property.

    \item
    \emph{Shuffling values of multiple attributes}:
    This operation extends the above by also shuffling subsets of values from $m$ additional attributes, randomly selected from the dataset. All selected attributes, including the confidential property-related one, are shuffled independently. This introduces variation in attribute joint distribution while preserving the confidential property.
    
    \item
    \emph{Mutating values of the confidential property-related attribute}:
    This operation randomly replaces a subset of values in the confidential property-related attribute with alternative valid values from its domain. Unlike shuffling, this operation directly changes the dataset’s confidential property.

    \item
    \emph{Mutating values of multiple attributes}:
    This operation extends the previous mutation by additionally replacing subsets of values in $m$ other randomly selected attributes. Consequently, this introduces variation into both the joint distribution of attributes and the confidential property itself.

\end{itemize}

Moreover, to ensure the perturbed dataset $D_k^{S^\prime}$ remains a realistic variant of the original dataset $D_k^S$, we constrain the extent of perturbation using a perturbation budget $\delta \in \mathbb{Z}^+$. Specifically, we require:
\begin{equation*}
|D_k^S - D_k^{S^\prime}|_0 \leq \delta,
\end{equation*}
where $|D_k^S - D_k^{S^\prime}|_0$ denotes the number of elements that differ between $D_k^S$ and $D_k^{S^\prime}$. This constraint ensures that each perturbed dataset remains close to its original counterpart, preserving the realism and distributional fidelity of perturbed datasets.

\vpara{Parameter approximations.} 
Given a dataset $D_k^{S^\prime}$ obtained by perturbing the original dataset $D_k^{S}$, we are now in a position to approximate the parameters $\theta_k^{\prime}$ of the shadow model that would have been trained on $D_k^{S^\prime}$. We denote such an approximated shadow model as $f_k^\prime$.
Concretely, instead of retraining $f_k^\prime$ on $D_k^{S^\prime}$, we estimate how the parameters $\theta_k$ of the reference shadow model $f_k^{\text{ref}}$, trained on $D_k^{S}$, would change due to the perturbation from $D_k^{S}$ to $D_k^{S^\prime}$. This allows us to obtain an approximation of $\theta_k^{\prime}$ efficiently.

Let $Z_k \subset D_k^{S}$ be the subset of samples that have at least one attribute value modified by the perturbation, and $Z_k^{\prime} \subset D_k^{S^{\prime}}$ represent the corresponding perturbed samples. The model parameter change is thus denoted as $\Delta(Z_k, Z_k^\prime)$, such that:
\begin{equation*}
    \theta_k^{\prime} = \theta_k + \Delta(Z_k, Z_k^\prime).
\end{equation*}
Following \citet{koh2017understanding} and \citet{WarneckePWR23}, we approximate the parameter change $\Delta(Z_k, Z_k^\prime)$ according to the theorem below.

\begin{theorem}
\label{theorem: approx}
Suppose the loss function $l(\cdot;\cdot)$ employed in shadow model training is twice differentiable and strictly convex. Then, the shadow model parameter change due to dataset perturbation can be approximated as: 
\begin{equation}\label{eq:est_para}
\Delta(Z_k, Z_k^\prime)\approx-\frac{1}{|D_k^S|}H_{\theta_k}^{-1}(\underset{z^\prime\in Z_k^\prime}{\sum}\nabla_{\theta_k} l(\theta_k;z^\prime)-\underset{z \in Z_k}{\sum}\nabla_{\theta_k} l(\theta_k;z)),
\end{equation}
where $|D_k^S|$ is the size of training dataset, $H_{\theta_k}^{-1}$ denotes the inverse Hessian of the empirical loss $\frac{1}{|D_k^S|}\sum_{i=1}^{|D_k^S|} l(f_k^{\text{ref}}(X_i^S),Y_i^S)$ evaluated at $\theta_k$, and $\nabla_{\theta_k} l(\theta_k;z)$ is the gradient of the loss with respect to model parameters $\theta_k$ for sample $z$.
\end{theorem}

Proof: See Appendix \ref{appendix: proof_approx}.

Algorithm~\ref{alg:effi} summarizes our  approximation strategy. By doing so, we eliminate the need to train all $N$ shadow models from scratch. Instead, we train only $K$ reference shadow models (lines \ref{alg_line:effi:ref_start}-\ref{alg_line:effi:ref_end}) and apply the proposed parameter approximation strategy to the remaining $N - K$ models (lines \ref{alg_line:effi:app_start}-\ref{alg_line:effi:app_end}). This approach significantly reduces the overall computational overhead compared to conventional methods that require independent training of all $N$ shadow models.  Consequently, our strategy substantially improves the computational efficiency of the proposed defense method and ensures its practical viability for real-world secure model-sharing scenarios.

\begin{algorithm}[t]
\caption{Approximation Strategy for Shadow Model Training} 
\label{alg:effi}
\begin{flushleft}
\textbf{Input:} Auxiliary dataset $D^{\text{adv}}$; total number of shadow models $N$; number of reference shadow models $K$. \\
\vspace*{-0.1in}\textbf{Output:} Shadow models $f^S_i$ with parameters $\theta_i$ and corresponding training datasets $D^S_i$, for $i = 1, \dots, N$.
\end{flushleft}
\begin{algorithmic}[1]
\FOR{$i \in [1, K]$} \label{alg_line:effi:ref_start}
\vspace*{-0.1in}    \STATE Sample training dataset $D^S_i$ from $D^{\text{adv}}$.
\vspace*{-0.1in}    \STATE Train reference shadow model $f_i^{\text{ref}}$ with parameters $\theta_i = \arg\min_\theta \mathcal{L}_{\mathcal{T}}(\theta; D^S_i)$.
\vspace*{-0.1in}    \STATE Set $f^S_i \leftarrow f_i^{\text{ref}}$.
\vspace*{-0.1in}\ENDFOR \label{alg_line:effi:ref_end}
\vspace*{-0.1in}\FOR{$i \in [K+1, N]$} \label{alg_line:effi:app_start}
\vspace*{-0.1in}    \STATE Uniformly sample $k \sim [1, K]$.
\vspace*{-0.1in}    \STATE Generate perturbed dataset $D^{S^\prime}_k$ from $D^S_k$ using the procedure in Eq~\eqref{eq:d_per}.
\vspace*{-0.1in}    \STATE Apply Theorem~\ref{theorem: approx} to approximate  parameters $\theta_k^\prime$ of shadow model $f_k^\prime$ if trained on $D_k^{S^\prime}$.
\vspace*{-0.1in}    \STATE Set $f^S_i \leftarrow f_k^\prime$ and $\theta_i \leftarrow \theta_k^\prime$.
\vspace*{-0.1in}    \STATE Set $D^S_i \leftarrow D^{S^\prime}_k$.
\vspace*{-0.1in}\ENDFOR \label{alg_line:effi:app_end}
\vspace*{-0.1in}\RETURN Shadow models $f^S_i$ with parameters $\theta_i$ and their training datasets $D^S_i$, for $i = 1, \dots, N$.
\end{algorithmic}
\end{algorithm}

\subsection{Overall Algorithm}
\label{subsec:overall alg}

Given a vanilla target ML model $f_{\theta^{(0)}}$ trained on the model provider's proprietary dataset $D$ with a confidential property $P$, the overall defense procedure of $\text{D}$-$\text{S}^2$HARE for producing a secure target model is summarized in Algorithm~\ref{alg:Dshare}. 

Specifically, line~\ref{alg_line:Dshare:alg_effi} of Algorithm~\ref{alg:Dshare} invokes the proposed approximation strategy (Algorithm~\ref{alg:effi}) to efficiently acquire $N$ shadow models, thereby significantly reducing the computational overhead associated with shadow model training.  Using these shadow models,  lines~\ref{alg_line:Dshare:att_data_start}-\ref{alg_line:Dshare:att_data_end} derive each model’s information $\mathcal{F}^S_i$ and compute the corresponding training dataset’s confidential property $ P^S_i$; together, these form the dataset $\{(\mathcal{F}^S_i, P^S_i)\}_{i=1}^N$ for training Responsive CPI attack models that simulate real-world adversaries. 
Line~\ref{alg_line:Dshare:att_initial} initializes a Responsive CPI attack model that is specifically tailored to model information $\mathcal{F}_{\theta^{(0)}}$ of the vanilla target ML model. Starting from line~\ref{alg_line:Dshare:ar_start}, the attack-defense arms race proceeds iteratively: the model provider first updates the target model parameters to reduce the attack’s effectiveness while preserving utility (line~\ref{alg_line:Dshare:ar_target}), after which the simulated adversary refines its attack model in response (line~\ref{alg_line:Dshare:ar_att}). The process repeats until convergence or until the maximum iteration count $T$ is reached, ultimately yielding a secure target model $f_{\theta^*}$ that is robust to Responsive CPI attacks yet retains high utility for its original task. In addition, a computational complexity analysis of $\text{D}$-$\text{S}^2$HARE can be found in Appendix~\ref{appendix:complexity}.

\begin{algorithm}[t]
\caption{{Overall Defense Procedure of $\text{D}$-$\text{S}^2$HARE}} 
\label{alg:Dshare}
\begin{flushleft}
\textbf{Input:} Vanilla target ML model $f_{\theta^{(0)}}$ and accessible model information $\mathcal{F}_{\theta^{(0)}}$;\\
\vspace*{-0.1in}\hspace*{0.42in} proprietary dataset $D$ and its confidential property $P$;\\
\vspace*{-0.1in}\hspace*{0.42in} convergence threshold $\epsilon$; maximum iterations $T$; trade-off hyperparameter $\lambda$.\\
\vspace*{-0.07in} \textbf{Output:} Secure target ML model $f_{\theta^*}$.
\end{flushleft}
\begin{algorithmic}[1]
\vspace*{-0.1in}\STATE Acquire shadow model $f^S_i$ with parameters $\theta_i$ and training dataset $D^S_i$ by Algorithm~\ref{alg:effi}, for $i=1,\dots,N$. \label{alg_line:Dshare:alg_effi}
\vspace*{-0.1in}\FOR{$i\in[1,N]$} \label{alg_line:Dshare:att_data_start}
\vspace*{-0.1in}    \STATE Derive model information $\mathcal{F}^S_i$ of shadow model $f^S_i$.
\vspace*{-0.1in}    \STATE Compute confidential property $P^S_i$ of dataset $D^S_i$.
\vspace*{-0.1in}\ENDFOR
\vspace*{-0.1in}\STATE Construct training dataset $\{(\mathcal{F}^S_i, P^S_i)\}_{i=1}^N$ for the attack model. \label{alg_line:Dshare:att_data_end}
\vspace*{-0.1in}\STATE Set iteration count $t=0$.
\vspace*{-0.1in}\STATE Initialize Responsive CPI attack $h_{\phi^{(1)}}$ with  $\phi^{(1)}=\arg\min_\phi \mathcal{L}_{\mathcal{A}}(\phi,\mathcal{F}_{\theta^{(0)}})$. \hfill $\triangleright$ Eq.~\eqref{eq:La final} \label{alg_line:Dshare:att_initial}
\vspace*{-0.1in}\REPEAT \label{alg_line:Dshare:ar_start}
\vspace*{-0.1in}\STATE $t \leftarrow  t+1$.
\vspace*{-0.1in}\STATE Update target ML model to $f_{\theta^{(t)}}$ with $\theta^{(t)} =\arg\max_{\theta}  \left[\mathcal{L}_{\mathcal{P}}(P, h_{\phi^{(t)}}(\mathcal{F}_\theta)) - \lambda\mathcal{L}_{\mathcal{T}}(\theta;D)\right]$ .\hfill $\triangleright$ Eq.~\eqref{eq:Lt} \label{alg_line:Dshare:ar_target}
\vspace*{-0.1in}\STATE Refine Responsive CPI attack model to $h_{\phi^{(t+1)}}$ with $\phi^{(t+1)}=\arg\min_\phi \mathcal{L}_{\mathcal{A}}(\phi,\mathcal{F}_{\theta^{(t)}})$. \hfill $\triangleright$ Eq.~\eqref{eq:La final} \label{alg_line:Dshare:ar_att}
\vspace*{-0.1in} \UNTIL{$|\theta^{(t)}-\theta^{(t-1)}|<\epsilon$ $\bigvee$ $t\ge T$.}
\vspace*{-0.1in}\STATE $f_{\theta^*}\leftarrow f_{\theta^{(t)}}$
\vspace*{-0.1in} \RETURN Secure target ML model $f_{\theta^*}$.
\end{algorithmic}
\end{algorithm}

\section{Empirical Evaluation}

This section provides a comprehensive evaluation of our proposed defense method.
We describe the experimental setup, including the model-sharing scenario, dataset, evaluation metrics,  and benchmark methods in Section \ref{subsec:setup}. Next, we report the evaluation results on CPI-attack robustness and utility retention for our proposed defense and benchmarks in Section \ref{subsec:performance_recpi}. We then present the robustness checks in Section~\ref{subsec:robustness} and the defense efficiency assessment in Section~\ref{subsec:defense_efficiency}. Lastly, we analyze the performance contribution of key components in our proposed method in Section \ref{subsec:ablation}. Additional experiments, such as parameter sensitivity analysis, are provided in Appendix~\ref{appendix:sensitivity}.

\subsection{Experimental Setup}
\label{subsec:setup}
\vpara{Scenario and dataset.}
We evaluated our method in a model-sharing scenario where a bank (the model provider) intends to share an ML telemarketing model (i.e., the target model).  The target task of this telemarketing model is to predict whether a customer will subscribe to a long-term deposit, to aid banks in targeting their telemarketing efforts toward customers most likely to respond. The model provider must protect the confidential property of the telemarketing model's training dataset from being inferred by potential adversaries, while ensuring that the telemarketing model effectively performs the target task. The model provider can share the target model in two ways: white-box and black-box settings. In the white-box setting, the model provider directly shares the parameters and architectures of the target model. In the black-box setting, the model provider shares only query access to the target model without exposing the model details. Our evaluation was conducted under both settings.

We utilized the Bank Marketing dataset \citep{misc_bank_marketing_222} to simulate this model-sharing scenario.
The dataset contains records related to telemarketing campaigns of a Portuguese banking institution. Each record describes a customer's information with three main categories:
(1) demographic attributes, including age, job, marital status and education level; (2) bank marketing actions, including contact channels via cellular or telephone, last contact time, number of contacts during and before the campaign, and the outcome of the previous marketing campaign; (3) banking profile, including whether the customer has a personal loan, a housing loan, the customer's average yearly balance, and whether the customer has defaulted.
Among these, attributes in the banking profile are particularly confidential, as their properties reflect key indicators of a bank’s risk management and overall asset quality. Here, we designated the default rate, i.e., the percentage of customers who have defaulted, as the confidential property of the dataset. We also defined an additional confidential property for robustness checks in Section \ref{subsec:robustness}.

The Bank Marketing dataset contains 45,211 records.
We randomly sampled 25,000 records from the dataset for the model provider to constitute the provider's confidential data.
The remaining records were the adversary's auxiliary dataset. 
To comprehensively evaluate performance under varying cases where the model provider and adversary possess distinct datasets, we repeated this sampling process 100 times. This resulted in 100 unique cases,  each with a different partitioning of the model provider’s private data and the adversary’s auxiliary dataset.

\vpara{Evaluation metrics.}
The performance of each method was evaluated using two metrics, \textit{success rate} of CPI attacks and \textit{target accuracy}, to reflect the CPI-attack robustness and utility retention capacity of the shared target model, respectively.

Specifically, to evaluate the robustness of defense methods against CPI attacks, we framed the CPI attack task as a binary classification task that predicts whether the dataset’s default rate is below or above 5\%. Additional CPI attack tasks were examined as part of the robustness checks in Section~\ref{subsec:robustness}.
Accordingly, the success rate of CPI attacks is defined as the proportion of shared target models for which the attack model accurately infers the confidential property. We have  
\begin{equation*}
 \text{Attack success rate} = \frac{N_p}{N_t},
\end{equation*}
where $N_p$ is the number of target models whose confidential properties are correctly inferred by the attack model, and $N_t$ is the total number of target models. A lower success rate indicates a stronger ability of defense methods to defend against CPI attacks.

The target accuracy was introduced to evaluate the performance of the target ML models in the target telemarketing task. 
It is calculated as the ratio of the correctly predicted customer outcomes ($N_m$) to the total number of customers ($N_c$):
\begin{equation*}
 \text{Target accuracy} = \frac{N_m}{N_c}.
\end{equation*}
Higher target accuracy suggests that the target model maintains its utility effectively.

\vpara{Benchmark methods.}
We compared our defense method $\text{D}$-$\text{S}^2$HARE with the following benchmarks. 
\begin{itemize}
    \item  W/O defense, which is the simplest baseline to train the target model as usual, without any defenses. Compared with this benchmark, we can assess the robustness of defense methods against CPI attacks.  
\end{itemize}

In addition, we compared our method with existing noise-based defense methods, which introduce noise into different phases during the target model training. The comparison with these methods demonstrates the advantage of model-based defense methods in maintaining utility.
\begin{itemize}
    \item  Noisy Label \citep{ganju2018property}, which introduces noise to the labels of the target task by flipping a certain proportion of labels.
    \item  DP-SGD \citep{mahloujifar2022property}, which adds Laplace noise to the target model gradients to obscure sensitive information during training.
    \item  Resampling \citep{suri2023dissecting}, 
    which under-samples the training data for the target model so that the confidential property is altered.
\end{itemize}

Moreover, we benchmarked our method against state-of-the-art model-based defense methods, which aim to defend against a presumed adversary.  This comparison highlights the superiority of our method over existing model-based defenses.
\begin{itemize}
    \item Property Unlearning \citep{stock2022lessons}, which modifies the parameters of the target model to reduce the efficacy of the CPI attack model proposed by \citet{ganju2018property}. This method is specifically designed for the white-box setting only.

    \item Adversarial Defense \citep{stockproperty}, which adjusts the parameters of the target model to minimize the effectiveness of the CPI attack proposed by \citet{ganju2018property}, while simultaneously preserving the utility of the target ML model. 
\end{itemize}

\vpara{Implementation details.}
In our model-sharing scenario, we assumed the target model is a Multilayer Perceptron (MLP) with 2 hidden layers of sizes 32 and 16. For the model provider, we sampled 10,000 records for training the target ML model and the remaining 15,000 records for simulating CPI attacks.
When implementing our defense method, we set the batch size to 32 and the learning rate to 1e-3. Both $\mathcal{L}_\mathcal{P}$ and $\mathcal{L}_\mathcal{T}$ in the objective function~\eqref{eq:obj_def} were instantiated as cross-entropy losses. The default value of the trade-off parameter $\lambda$ was set as 0.3. 
When simulating the adversary using the Responsive CPI attack, we trained $K=100$ reference shadow models with the same architecture as the target model. From each reference shadow model's training dataset, we generated 4 perturbed datasets for approximated shadow models, obtaining 400 approximated shadow models. 
When implementing the Responsive CPI attack in the black-box setting, we randomly sampled 1000 samples from the adversary's auxiliary data as the query set. This query set was then used to obtain outputs from both the shadow models and the target model, enabling the attack model to learn and infer the confidential properties based on these outputs.
In the approximation strategy, we set the perturbation budget $\delta$ at 1,000.
The attack model was an MLP network with 3 hidden layers of sizes 32, 16 and 8. The $l(\cdot,\cdot)$ term used in the attack loss $\mathcal{L}_\mathcal{A}$  was instantiated as the cross-entropy loss. When estimating the training sample weight $r_i$, we set the kernel width $\sigma^2$ to 0.75. 

For the benchmark methods, we followed \citet{ganju2018property} to implement Noisy Label by flipping labels for the target task. DP-SGD was implemented following \citet{opacus}. We adapted the publicly released codes\footnote{\href{https://github.com/iamgroot42/propertyinference}{https://github.com/iamgroot42/propertyinference}} from \citet{suri2023dissecting} for implementing Resampling and Property Unlearning. Additionally, we implemented Adversarial Defense based on the algorithm described by \citet{stockproperty}. All parameters were set to their default values, except for the confidentiality-utility trade-off parameters, if such parameters exist for a given benchmark. Specifically, 
the confidentiality-utility trade-off parameter is typically used to control the amount of allowed confidential information leakage. Generally speaking, as the allowed confidential information leakage increases, the model's robustness to CPI attacks decreases, while its utility (i.e., target accuracy) improves. Except for Property Unlearning, all benchmarks include this trade-off parameter. To ensure comparability across benchmarks, we conducted a comprehensive analysis of performance under different trade-off parameter settings. Details on the settings of trade-off parameters for each compared method are provided in Appendix~\ref{appendix:implementation}.

\subsection{Performance Under Responsive CPI Attack}
\label{subsec:performance_recpi}
In this section, we evaluated the performance of our proposed defense method and other benchmarks against the Responsive CPI attack, the most realistic and strongest attack that captures the adaptive nature of real-world adversaries. Performance of the defenses under other non-responsive CPI attacks is reported in Section~\ref{subsec:robustness}. Specifically, the performance of each defense method was evaluated using two metrics: the success rate of the CPI attack, which evaluates the robustness of the target ML model, and the target accuracy, which reflects the model's utility retention capability.

\begin{figure}[!h]
	\centering
	\subfloat[\centering $P$ as binary categories of default rate, white-box setting]{
		\begin{minipage}[t]{0.5\linewidth}
			\centering
			\includegraphics[width=2in]{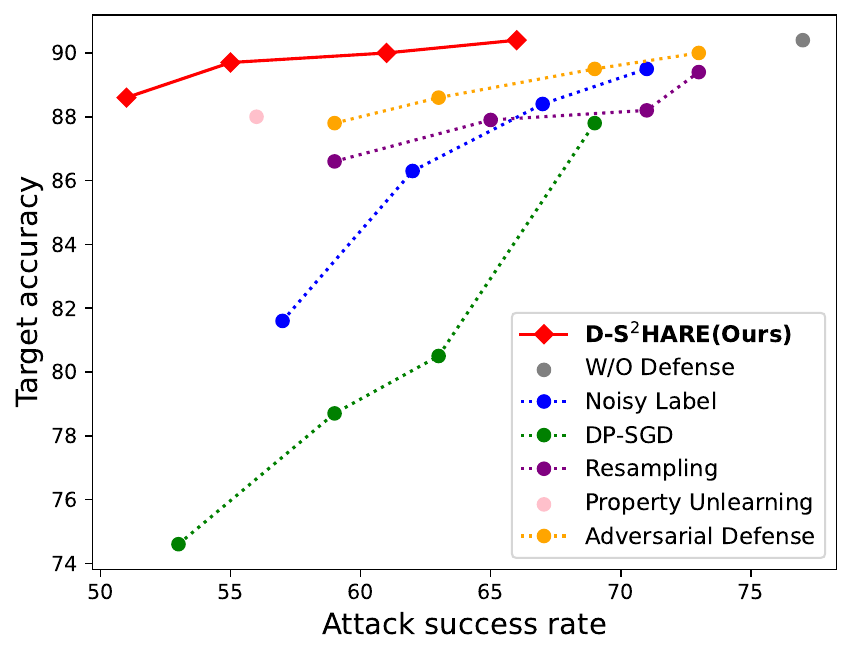}
		\end{minipage}
	}%
	\subfloat[\centering $P$ as binary categories of default rate, black-box setting]{
		\begin{minipage}[t]{0.5\linewidth}
			\centering
			\includegraphics[width=2in]{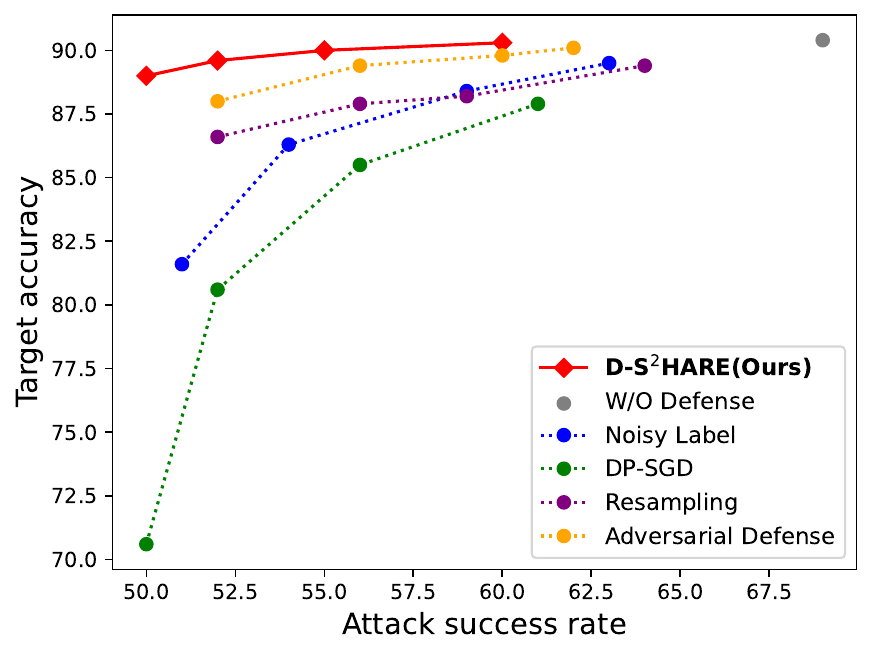}
		\end{minipage}
	}%
	\centering
    \caption{Performance Comparison between Our Method and Benchmarks under Varying Trade-off Parameters
    }
    \vspace{-1cm}
	\label{fig:main}
\end{figure}

Figure~\ref{fig:main} presents a comparison of our defense method and benchmark methods under both white-box (i.e., adversaries have access to model parameters) and black-box settings (i.e., adversaries have query access only). The horizontal axis represents the success rate of the Responsive CPI attack, while the vertical axis indicates the target accuracy. A lower value on the horizontal axis reflects higher robustness to Responsive CPI attacks, whereas a higher value on the vertical axis corresponds to better utility retention.
Except for Property Unlearning, which does not include a confidentiality-utility trade-off parameter, we evaluated each defense method under four different trade-off parameter settings. For each method, each point represents the result under one configuration. Notably, points closer to the upper-left corner indicate better overall performance, combining both higher robustness and better utility retention.

As shown in the figure, compared to all the benchmark methods, our defense method positions in the upper-left corner under all settings, indicating the best performance in both CPI-attack robustness (lower attack success rate) and utility retention (higher target accuracy). By comparison,
noise-based defense methods (i.e., Noisy Label, DP-SGD and Resampling) often result in a significant reduction in model utility, which degrades the target task's performance and makes them less suitable for secure model-sharing.
Existing model-based defense methods (i.e., Property Unlearning and Adversarial Defense) typically achieve better utility but fall short in defending against Responsive CPI attack compared to our method.
The superior performance of our method is attributed to the consideration of responsive nature of CPI attacks during the training process of the target model. By progressively defending against responsive attacks, our method enhances the robustness against CPI attacks.

\subsection{Robustness Check}
\label{subsec:robustness}
Moreover, we conducted a series of robustness checks by varying the experimental settings to assess whether our method consistently outperforms others. 
Specifically, we examined performance of the compared methods under four types of robustness settings.

\begin{figure}[!h]
\centering
\vspace{-0.5cm}
	\subfloat[\centering $P$ as 3-class categories of default rate, white-box]{
		\begin{minipage}[t]{0.5\linewidth}
			\centering
			\includegraphics[width=2in]{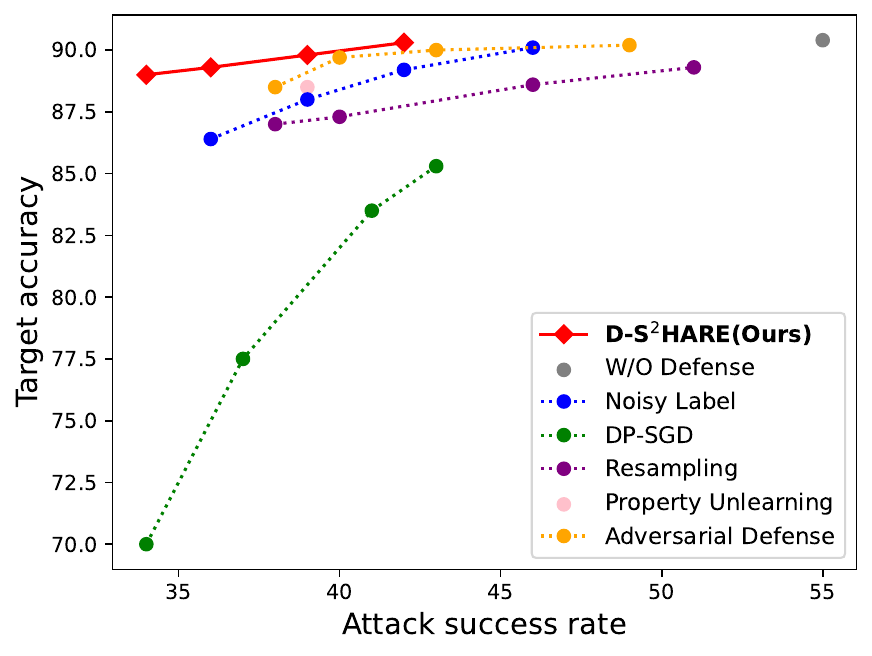}
		\end{minipage}
        \label{fig:task_w}
	}%
	\subfloat[\centering $P$ as 3-class categories of default rate, black-box]{
		\begin{minipage}[t]{0.5\linewidth}
			\centering
			\includegraphics[width=2in]{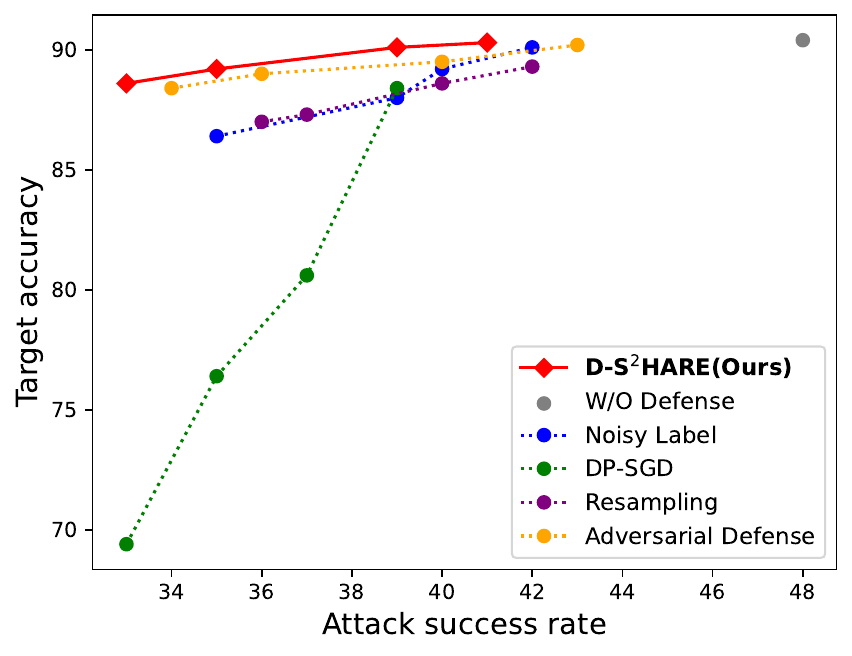}
		\end{minipage}
        \label{fig:task_b}
	}%
    
 	\subfloat[\centering $P$ as binary categories of housing loan rate, white-box]{
		\begin{minipage}[t]{0.5\linewidth}
			\centering
			\includegraphics[width=2in]{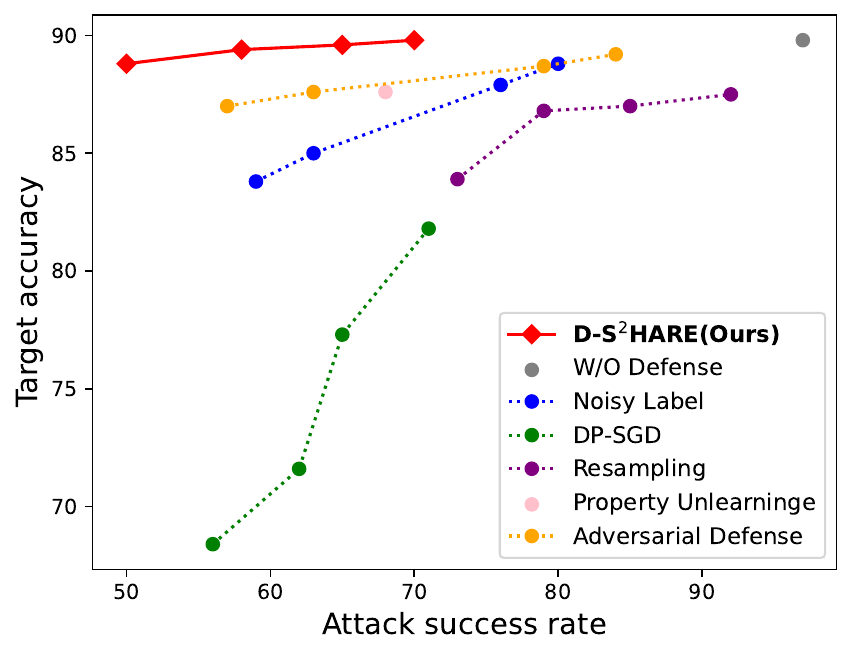}
		\end{minipage}
        \label{fig:p_w}
	}%
 	\subfloat[\centering $P$ as binary categories of housing loan rate, black-box]{
		\begin{minipage}[t]{0.5\linewidth}
			\centering
			\includegraphics[width=2in]{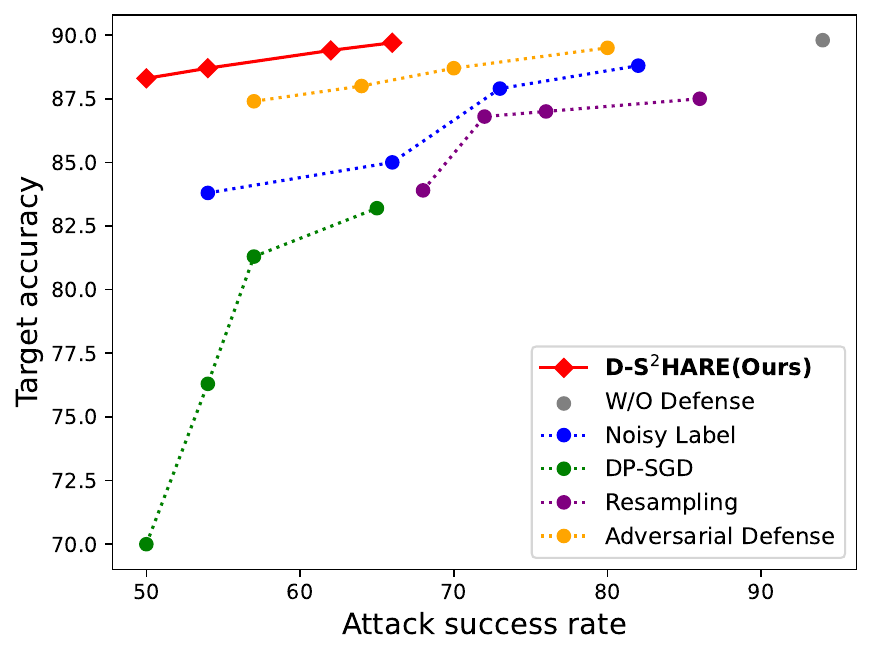}
		\end{minipage}
        \label{fig:p_b}
	}%
 
	\subfloat[\centering $P$ as binary categories of default rate, white-box, under non-responsive CPI attack proposed by \citet{ganju2018property}]{
		\begin{minipage}[t]{0.5\linewidth}
			\centering
			\includegraphics[width=2in]{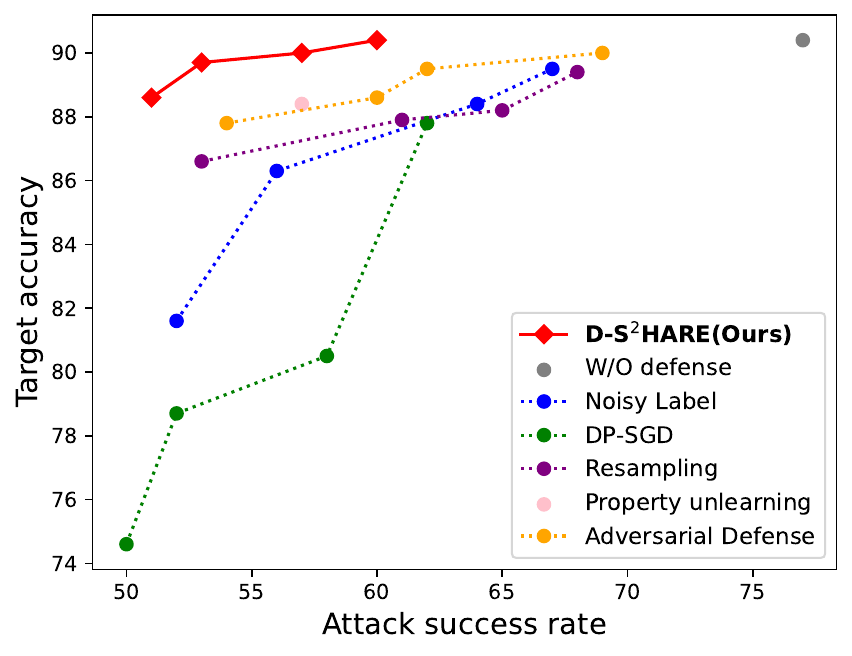}
		\end{minipage}
        \label{fig:nonre_w}
	}%
	\subfloat[\centering $P$ as binary categories of default rate, black-box, under non-responsive CPI attack proposed by \citet{zhang2021leakage}]{
		\begin{minipage}[t]{0.5\linewidth}
			\centering
			\includegraphics[width=2in]{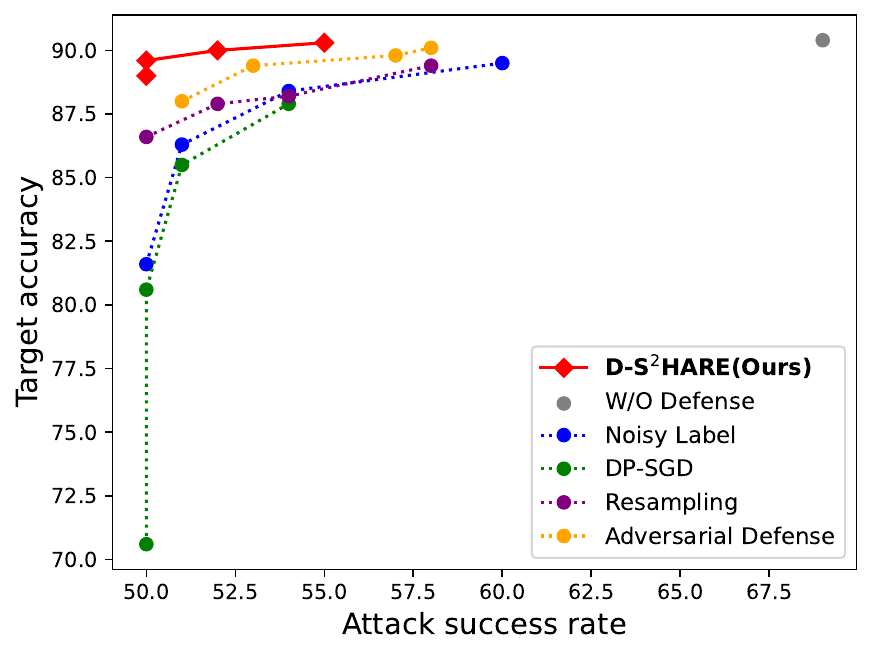}
		\end{minipage}
        \label{fig:nonre_b}
	}%
    
 	\subfloat[\centering $P$ as binary categories of default rate, white-box, logistic regression model as the target model]{
		\begin{minipage}[t]{0.5\linewidth}
			\centering
			\includegraphics[width=2in]{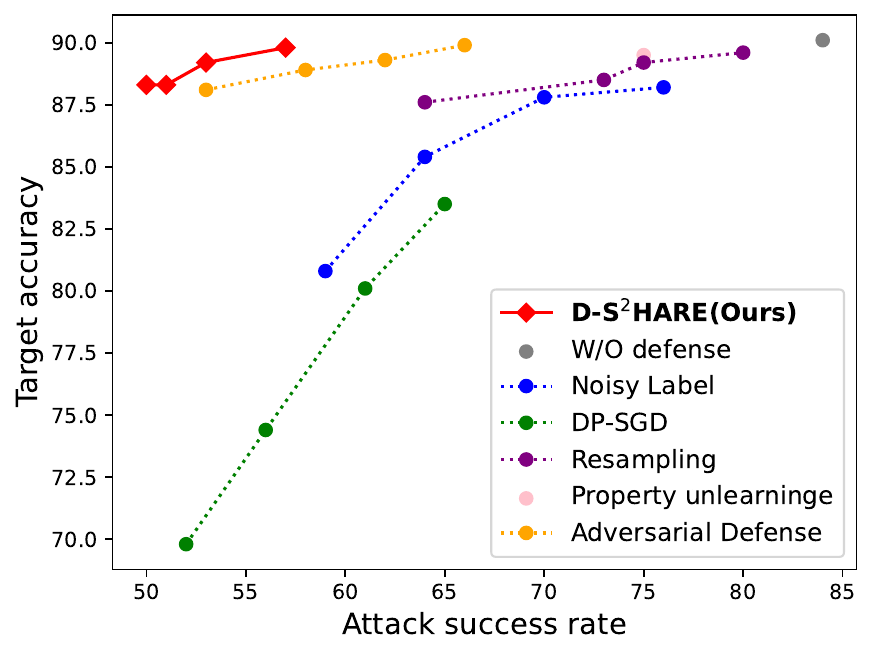}
		\end{minipage}
        \label{fig:target_w}
	}%
 	\subfloat[\centering $P$ as binary categories of default rate, black-box, logistic regression model as the target model]{
		\begin{minipage}[t]{0.5\linewidth}
			\centering
			\includegraphics[width=2in]{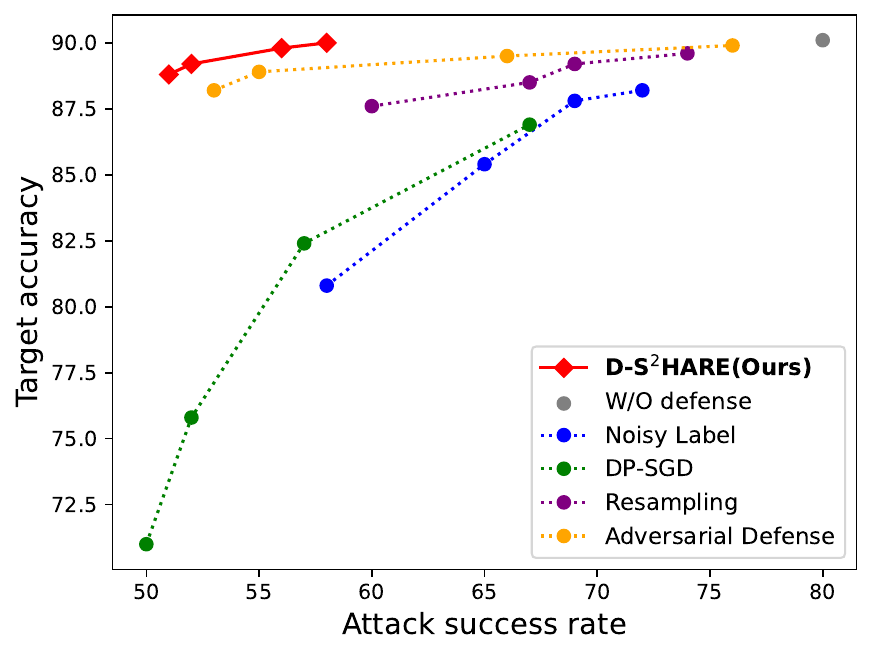}
		\end{minipage}
        \label{fig:target_b}
	}%
	\centering
	\caption{Robustness Checks
    }
    \vspace{-0.8cm}
	\label{fig:robustness}
\end{figure}

First, we altered the CPI attack task. Instead of the binary classification of the default rate, we further constructed a 3-class classification task that categorizes the default rate into one of three intervals: below 5\%, between 5\% and 10\%, or between 10\% and 15\%. The results are presented in Figure~\ref{fig:robustness}(a) and~\ref{fig:robustness}(b). 

Second, we evaluated performance of the methods under a different confidential property: the housing loan rate, defined as the mean value of the attribute indicating whether a customer has a housing loan. The corresponding CPI task here is a binary classification that determines whether it is below or above 50\%. The results are reported in Figure~\ref{fig:robustness}(c) and~\ref{fig:robustness}(d). 

Third, we examined the performance of our method and benchmark methods against non-responsive CPI attacks.
For white-box CPI attacks, we evaluated the performance of the methods against the attack proposed by \citet{ganju2018property}, which assumes that adversary can obtain the parameters of the target model and thus inferring the confidential property by sorted network neurons.
For black-box CPI attacks, we evaluated the performance of the methods against the attack proposed by \citet{zhang2021leakage}, which uses probability vectors from a specific set of queries to the target model to infer confidential properties. The results are shown in Figure~\ref{fig:robustness}(e) and~\ref{fig:robustness}(f).

Lastly, we replaced the target ML model with a logistic regression model to evaluate the robustness and superiority of our proposed defense across different model architectures.
The results are presented in Figure~\ref{fig:robustness}(g) and~\ref{fig:robustness}(h). 
Across all robustness checks, we can observe that our defense method consistently remains positioned at the upper-left corner of each figure, indicating superior performance in both CPI-attack robustness and utility retention under a wide range of experimental settings. We provide the implementation details of these robustness checks in Appendix \ref{appendix:implementation}.

\subsection{Defense Efficiency}
\label{subsec:defense_efficiency}
Table~\ref{tab:run_time} reports the training time of our method and the most competitive benchmarks, specifically existing model-based defense methods. Additionally, since no current methods account for the responsive nature of adversaries, a straightforward solution is to combine the best-performing benchmark, Adversarial Defense, with our proposed Responsive CPI attack. This integration initiates the arms race attack–defense framework to produce a secure target ML model. We refer to this modified version as Adversarial Defense+.

\begin{table}[!h]
\centering
\caption{ Training Time Comparison between $\text{D}$-$\text{S}^2$HARE and Existing Model-based Defenses (in Second)}
\label{tab:run_time}
\renewcommand{\arraystretch}{0.5}
\begin{tabular}{ c | cc | cc }
\Xhline{1pt}
 \multirow{2}{*}{\begin{tabular}[c]{@{}c@{}}Method\end{tabular}} & \multicolumn{2}{c|}{White-box} & \multicolumn{2}{c}{Black-box} \\
 & Training time & Time reduced & Training time & Time reduced \\ \hline
 \textbf{\begin{tabular}[c]{@{}c@{}}$\text{D}$-$\text{S}^2$HARE\\ (Ours)  \end{tabular}} & 141.86s & - & 258.81s & - \\
 Property Unlearning &  581.73s & 75.61\% & - & - \\
 Adversarial Defense & 566.31s & 74.95\% & 606.84s & 57.35\% \\
 Adversarial Defense+
 & 2641.57s & 94.63\% & 3147.39s & 91.78\% \\
\Xhline{1pt}
\end{tabular}
\end{table}
\vspace{-10pt}

The results show that our method is the most time-efficient,  achieving a substantial reduction in training time---specifically, by 74.95\% to 94.63\% in the white-box setting and by 57.35\% to 91.78\% in the black-box setting. The significant improvement in efficiency is attributed to our proposed approximation strategy, which eliminates the need for extensive training shadow models from scratch to simulate CPI attacks.
In comparison, the substantial computational cost of Adversarial Defense+ underscores that existing model-based defense methods, when accounting for responsive adversaries, exhibit very high training time, thereby failing to meet practical efficiency requirements.
Other model-based defense benchmarks exhibit longer runtime than ours, as they necessitate training all the shadow models from scratch when simulating CPI attacks.
In summary, our method not only achieves superior efficiency but also better performance in defending against CPI attacks and maintaining utility.

\subsection{Ablation Analysis}
\label{subsec:ablation}
We further conducted an ablation analysis to investigate the contribution of each component of our proposed defense method $\text{D}$-$\text{S}^2$HARE to its performance. Our focus is on two key methodological components: the consideration of Responsive CPI attacks and the approximation strategy.
More concretely, we first replaced the simulated Responsive CPI attack in our $\text{D}$-$\text{S}^2$HARE with existing CPI attacks, designating the resulting method as Ours-R. In the white-box setting, we replaced the Responsive CPI attack with the attack proposed by \citet{ganju2018property}, while in the black-box setting, we replaced it with the attack proposed by \citet{zhang2021leakage}. 
The performance difference between $\text{D}$-$\text{S}^2$HARE and Ours-R reveals the contribution of considering the responsive nature of CPI attacks.
Furthermore, to evaluate the contribution of the approximation strategy, we removed this component from our method and trained all shadow models from scratch. We designate this resulting method as Ours-A. The performance difference between $\text{D}$-$\text{S}^2$HARE and Ours-A uncovers the contribution of the approximation strategy.

\begin{table}[H]
\centering
\caption{Ablation Analysis of $\text{D}$-$\text{S}^2$HARE}
\label{tab:ablation}
\renewcommand{\arraystretch}{0.5}
\resizebox{.75\textwidth}{!}
{
\begin{tabular}{c | ccc | ccc @{}}
\Xhline{1pt}
\multicolumn{1}{c|}{{\begin{tabular}[c|]{@{}c@{}}  \end{tabular}}} &
 \multicolumn{3}{c |}{\begin{tabular}[c|]{@{}l@{}} White-box \end{tabular}} &
  \multicolumn{3}{c }{\begin{tabular}[c]{@{}l@{}} Black-box \end{tabular}}
  \\ & Success rate & Target acc & Run time  & Success rate &  Target acc & Run time   \\ \hline
\textbf{\begin{tabular}[c]{@{}c@{}}$\text{D}$-$\text{S}^2$HARE\\ (Ours)  \end{tabular}} & \textbf{0.55} & \textbf{0.897} & \textbf{141.86s} & \textbf{0.52} & \textbf{0.896}& \textbf{258.81s}\\
Ours-R & 0.60 & 0.900 & 135.71s & 0.61 & 0.890 & 232.03s  \\
Ours-A & 0.50 & 0.890 & 544.12s & 0.50 & 0.893 & 649.85s \\
\Xhline{1pt}
\end{tabular}}
\end{table}

Table \ref{tab:ablation} shows the performance of $\text{D}$-$\text{S}^2$HARE, Ours-R and Ours-A under both white-box and black-box settings. 
As shown, the attack success rate of Ours-R is notably higher than that of $\text{D}$-$\text{S}^2$HARE, highlighting the importance of considering responsive nature of CPI attacks during the defense process. Without this component, the shared target ML model fails to effectively defend against Responsive CPI attacks, resulting in reduced CPI-attack robustness. 
The variant Ours-A performs comparable to $\text{D}$-$\text{S}^2$HARE in terms of CPI-attack robustness and utility retention, but the training time is significantly higher. This underscores the necessity to improve efficiency through the proposed approximation strategy.
Overall, the ablation analysis highlights the substantial contributions of both the consideration of the responsive nature of adversaries and the approximation strategy to the effectiveness and efficiency of our defense method.

\section{Conclusion and Discussion}
\label{sec:con}
\subsection{Summary and Contributions}
\label{subsec:contrib}
Model sharing offers considerable business value, but concerns over data confidentiality remain a critical barrier. Such concerns arise from well-known CPI attacks, which can infer training dataset's confidential properties  solely from access to the shared model.
Although prior studies have proposed defenses against CPI attacks, existing methods often assume that attacks are static and overlook the responsive nature of real-world adversaries, making these defenses ineffective in practical model-sharing scenarios.
To address this limitation, we propose a novel defense method that explicitly accounts for the responsive nature of real-world adversaries. Specifically, we first devise a Responsive CPI attack that simulates such adversaries.  Building on this, we introduce an attack–defense arms race framework, where the target ML model and the attack model are iteratively enhanced. This iterative process results in a secure ML model that is robust against responsive CPI attacks. Furthermore, we propose and integrate a novel approximation strategy into our defense, which addresses the critical computational bottleneck of existing methods, thereby improving defense efficiency. Through extensive empirical evaluations across various realistic model-sharing scenarios, we demonstrate that our method outperforms existing defenses by more effectively defending against CPI attacks, preserving ML model utility, and reducing computational overhead.

Our study contributes to the extant literature in two ways. First, our study belongs to the computational genre of design science research in the IS field, which emphasizes the development of computational methods to address important business or societal problems and aims to make significant methodological contributions \citep{abbasi2024pathways,rai2017editor,padmanabhan2022machine}. In this regard, we focus on the valuable model‑sharing business and propose a novel defense method to address a critical information privacy issue arising in its applications.
Specifically, the key innovations of our novel defense method---Responsive CPI attack, arms race-based defense framework, and approximation strategy for computational efficiency---constitute the core methodological contributions of this work. Second, our study contributes to the research field of information privacy, which has increasingly captured the attention of IS scholars due to its substantial social and business implications (e.g., \citealp{li2017anonymizing, xu2022guest}). By proposing a novel method that enables model providers to share ML models securely, our work addresses a critical information privacy challenge that discourages model providers from engaging in the model-sharing. In doing so, our study promotes technological equity and helps mitigate the risks associated with technological monopolies.

\subsection{Implications and Future Work}
\label{subsec:implication_fw}
Our study offers several managerial implications for stakeholders in the model-sharing business. First, by adopting our secure model-sharing method, model providers can monetize their AI intellectual property (i.e., well-established ML models) without compromising data confidentiality. This approach addresses information privacy concerns and provides a reliable source of secure revenue streams. 
Second, our proposed Responsive CPI attack equips model providers with advanced tools for proactive risk management and risk assessment. This enables them to identify and address potential vulnerabilities early so as to better prepare both for sharing their ML models externally and reinforcing their internal privacy frameworks. This proactive stance is crucial for effective AI governance, helping ensure compliance with regulatory standards and promoting secure, trustworthy collaborations across various industries \citep{berente2021managing}. 
Third, our method significantly empowers model users, especially SMEs, by providing them access to well-established ML models without the heavy investments typically required for data collection and computing infrastructure. This access supports the growth and development of SMEs, allowing them to utilize advanced ML models to enhance their decision-making capabilities and drive business expansion. Consequently, SMEs can accelerate their innovation, optimize their operations, and significantly strengthen their market position in the industry.

This study also carries implications for policymakers and regulatory authorities who aim to make policy to foster secure and responsible AI ecosystems. 
First, as ML models increasingly become strategic assets for firms, our study demonstrates that data confidentiality risks can arise solely from sharing trained ML models. This suggests that data privacy regulations may need to extend their scope to explicitly regulate model-sharing practices, ensuring that derived or inferred confidential information from ML models is also covered under information privacy compliance standards.
Second, our study highlights that traditional defense strategies, which are typically designed based on static adversary assumptions, fail to safeguard ML models against responsive attacks. These responsive attacks more accurately reflect real-world threats. Therefore, regulatory guidelines should also evolve beyond static risk assumptions and incorporate evaluation criteria that assess model robustness against responsive privacy attacks.

Our study opens up multiple directions for future research. First, adapting our secure model-sharing paradigm to dynamic, streaming data environments presents a valuable future direction. In real-world applications, the model provider's training data  and the confidential properties contained within it often change over time. Therefore, future studies could extend our approach to allow for the continuous updating of the shared model in response to incoming data, while ensuring confidentiality, by incorporating continual learning and online learning approaches~\citep{bottou2003large, lesort2020continual}.
Second, our secure model-sharing paradigm introduces an attack–defense arms race framework where both the attack model and the target model are iteratively updated to achieve robustness against Responsive CPI attacks. Future research could explore methods to accelerate convergence in this iterative process to further improve the efficiency of secure model sharing.

\SingleSpacedXII
\bibliographystyle{informs2014}
\bibliography{references}

\newpage
\DoubleSpacedXI
\begin{APPENDICES}
\section{Proof of Theorems}

\subsection{Proof of Theorem \ref{theorem: weighted_loss}}
\label{appendix: proof_weight}
Recalling the definition of $R(h_\phi)$ in Eq.~\eqref{eq:test_error}, we have
\begin{align*}
    R(h_\phi)&=E_{\mathcal{D}_T}[l(h_\phi(\mathcal{F_\theta}),P)] \\
    &=\int_{Pr_{\mathcal{D}_T}(\mathcal{F}_\theta)\neq 0} Pr_{\mathcal{D}_T}(\mathcal{F}_\theta)l(h_\phi(\mathcal{F}_\theta),P) \,d\mathcal{F}_\theta\\
    &=\int_{Pr_{\mathcal{D}_S}(\mathcal{F}_i^S)\neq 0}\frac{Pr_{\mathcal{D}_T}(\mathcal{F}_i^S)}{Pr_{\mathcal{D}_S}(\mathcal{F}_i^S)} Pr_{\mathcal{D}_S}(\mathcal{F}_i^S)l(h_\phi(\mathcal{F}_i^S),P_i^S)\,d\mathcal{F}_i^S\\
    &= \int_{Pr_{\mathcal{D}_S}(\mathcal{F}_i^S)\neq 0} r_i Pr_{\mathcal{D}_S}(\mathcal{F}_i^S)l(h_\phi(\mathcal{F}_i^S),P_i^S)\,d\mathcal{F}_i^S \\
    &=E_{\mathcal{D}_S}[r_i l(h_\phi(\mathcal{F}_i^S),P_i^S)] \\
    &=\frac{1}{N} \sum_{i=1}^N E_{ \mathcal{D}_S}[r_i l(h_\phi(\mathcal{F}_i^S),P_i^S)] \\
    &=E_{\mathcal{D}_S}[\frac{1}{N} \sum_{i=1}^N r_i l(h_\phi(\mathcal{F}_i^S),P_i^S)],
\end{align*}
where the third equation holds under the assumption that the support of $\mathcal{D}_T$ is contained in the support of $\mathcal{D}_S$, the fourth equation uses the definition $r_i=\frac{Pr_{\mathcal{D}_T}(\mathcal{F}_i^S)}{Pr_{\mathcal{D}_S}(\mathcal{F}_i^S)}$, and the last two equations hold under the assumption that the training samples $(\mathcal{F}_i^S,P_i^S)$ are drawn i.i.d..

\subsection{Proof of Theorem \ref{theorem: r_i}}
\label{appendix: proof_r_i}
Recall that we have $r_i = \frac{\text{Pr}_{\mathcal{D}_T}(\mathcal{F}_i^S)}{\text{Pr}_{\mathcal{D}_S}(\mathcal{F}_i^S)}$ that adjusts the contribution of each training sample based on the probability density of $\mathcal{F}_i^S$ under the training sample distribution $\mathcal{D}_S$ relative to that under the test distribution $\mathcal{D}_T$. 
A good estimate of $r_i$, denoted by $\hat{r}_i$, is thus supposed to ensure that 
\begin{equation*}
    \hat{r}_i \text{Pr}_{\mathcal{D}_S}(\mathcal{F}_i^S) \approx \text{Pr}_{\mathcal{D}_T}(\mathcal{F}_i^S), \forall i=1,\dots, N.
\end{equation*}
This can be achieved by minimizing the Kullback-Leibler (KL) divergence of the probability density under the test distribution $\mathcal{D}_T$ from the probability density under the adjusted training sample distribution with respect to $r_i$:
\begin{align*}
    \min_{r_i} \text{KL}[\text{Pr}_{\mathcal{D}_T}(\mathcal{F}_i^S)\|r_i\text{Pr}_{\mathcal{D}_S}(\mathcal{F}_i^S)], \forall i=1,\dots, N,
\end{align*}
where the KL divergence is given by
\begin{align*}
\text{KL}[\text{Pr}_{\mathcal{D}_T}(\mathcal{F}_i^S)\|r_i \text{Pr}_{\mathcal{D}_S}(\mathcal{F}_i^S)] &=\int_{\mathcal{D}_T} \text{Pr}_{\mathcal{D}_T}(\mathcal{F}_i^S)\log\frac{\text{Pr}_{\mathcal{D}_T}(\mathcal{F}_i^S)}{r_i \text{Pr}_{\mathcal{D}_S}(\mathcal{F}_i^S)} d\mathcal{F}_i^S \\
&=\int_{\mathcal{D}_T} \text{Pr}_{\mathcal{D}_T}(\mathcal{F}_i^S)\log\frac{\text{Pr}_{\mathcal{D}_T}(\mathcal{F}_i^S)}{\text{Pr}_{\mathcal{D}_S}(\mathcal{F}_i^S)} d\mathcal{F}_i^S - \int_{\mathcal{D}_T} \text{Pr}_{\mathcal{D}_T}(\mathcal{F}_i^S)\log r_i d\mathcal{F}_i^S.
\end{align*}
Since the first term in the last equation is independent of $r_i$, the optimization problem focusing on minimizing the KL divergence with respect to $r_i$ is equivalent to maximizing the second term:
\begin{equation*}
    \int_{\mathcal{D}_T} \text{Pr}_{\mathcal{D}_T}(\mathcal{F}_i^S)\log r_i d\mathcal{F}_i^S \approx \frac{1}{N_T}\sum_i^{N_T}\log r_i,
\end{equation*}
where $N_T$ is the number of samples from the test distribution (i.e., target models in our problem).

Note that $r_i$ is non-negative by definition. Furthermore, to ensure that the probability density under the adjusted training sample distribution $r_i \text{Pr}_{\mathcal{D}_S}(\mathcal{F}_i^S)$ forms a valid probability density function, $r_i$ must satisfy the normalization constraint:
\begin{equation*}
    1=\int_{\mathcal{D}_S}r_i \text{Pr}_{\mathcal{D}_S}(\mathcal{F}_i^S)d\mathcal{F}_i^S \approx \frac{1}{N}\sum_i^N r_i.
\end{equation*}
Consider the training sample weight $r_i$ modeled as a linear model as suggested by \citep{sugiyama2007direct}:
\begin{equation*}
    r_i = \sum_{j=1}^{N_T} \alpha_j K_\sigma (\mathcal{F}_i^S, \mathcal{F}_j),
\end{equation*}
where $K_\sigma (\mathcal{F}_i^S, \mathcal{F}_j)=exp\{-\frac{\|\mathcal{F}_i^S-\mathcal{F}_j\|^2}{2\sigma^2}\}$ is the Gaussian kernel with kernel width $\sigma$, $\mathcal{F}_i^S$ denotes the model information of the $i$-th training sample, $\mathcal{F}_j$ denotes the model information of the $j$-th target ML model, and $\{\alpha_j\}_{j=1}^{N_T}$ are the kernel weights to be determined.
Given the non-negativity and normalization constraints, 
the training sample weight $r_i$ can be estimated by solving the following optimization problem:
\begin{align*}
    & \min_{\{\alpha_j\}_{j=1}^{N_T}} \frac{1}{N_T}\sum_i^{N_T}\log \left( \sum_{j=1}^{N_T} \alpha_j K_\sigma (\mathcal{F}_i^S, \mathcal{F}_j)\right)\\
    & \text{subject to} \: \frac{1}{N}\sum_{i=1}^N r_i=1  \: \text{and} \: r_i\geq0, \forall i=1,\dots, N.
\end{align*}

In our model-sharing scenario, the model provider typically shares only one target model, implying that $N_T=1$. In this case, $r_i$ simplifies to $r_i=\alpha K_\sigma(\mathcal{F}_i^S, \mathcal{F}_\theta)$, where $\mathcal{F}_\theta$ is the model information of the target model to be shared.
The optimization problem then reduces to
\begin{align*}
\max_\alpha \: & \log(\alpha K_\sigma (\mathcal{F}_i^S, \mathcal{F}_\theta)) \\
 \text{subject to} \: &\frac{1}{N}\sum_i^N \alpha K_\sigma (\mathcal{F}_i^S, \mathcal{F}_\theta)=1 \: \text{and} \: \alpha \geq 0.
\end{align*}
The closed-form solution of this problem is obtained by solving for $\alpha$:
\begin{equation*}
    \alpha=\frac{N}{\sum_{j=1}^N K_\sigma (\mathcal{F}_j^S, \mathcal{F}_\theta)}.
\end{equation*}
Substituting this into $r_i = \alpha K_\sigma (\mathcal{F}_i^S, \mathcal{F}_\theta)$, we obtain the estimate of training sample weight $r_i$ as
\begin{equation*}
    \hat{r}_i=\frac{N\cdot K_\sigma (\mathcal{F}_i^S, \mathcal{F}_\theta)}{\sum_{j=1}^N K_\sigma (\mathcal{F}_j^S, \mathcal{F}_\theta)}.
\end{equation*}

\subsection{Theorem \ref{theorem:convergence} And Its Proof}
\label{appendix:proof_convergence}
We first define the necessary notations to ensure clarity and simplify the convergence analysis. Subsequently, we outline the assumptions required for the convergence analysis. Following this, we detail Theorem \ref{theorem:convergence} and finally provide its proof.

\vpara{Notations.}
Let $\mathcal{L}_{\text{defend}}(\theta):= \lambda\mathcal{L}_\mathcal{T}(\theta)-\mathcal{L}_\mathcal{P}(\phi^*(\theta),\mathcal{F}_\theta)$ be the defender's objective function, where $\mathcal{L}_\mathcal{T}(\theta)$ is the average loss $l(f_\theta(X_i), Y_i))$ over the target model's training dataset, and $\mathcal{L}_\mathcal{P}(\phi^*(\theta),\mathcal{F}_\theta)=l(h_{\phi^*}(\mathcal{F}_\theta), P))$. Thus, 
we define the defender's per-sample loss as $l_{\text{defend}} (\theta, \phi^*(\theta))=\lambda l(f_\theta(X_i), Y_i))-l(h_{\phi^*}(\mathcal{F}_\theta), P))$, with $\mathcal{L}_{\text{defend}}(\theta)=E[l_{\text{defend}}(\theta,\phi^*(\theta))]$. Similarly, the adversary's per-sample loss is given by $l_{\text{adv}}(\theta, \phi)=r_i l(h_\phi(\mathcal{F}^S_i), P^S_i)$ for each training sample $(\mathcal{F}^S_i, P^S_i)$. 
Let $h_{\text{a}}^t$ denote the unbiased stochastic gradient of $l_{\text{adv}}(\theta^{(t)}, \phi^{(t)})$, and let $h_{\text{d}}^t$ denote the approximate stochastic gradient of $\mathcal{L}_{\text{defend}}(\theta^{(t)})$ at the $t$-th iteration.
We define $\overline{\nabla} l_{\text{defend}}(\theta, \phi) := \nabla_\theta l_{\text{defend}}(\theta, \phi)-\nabla^2_{\theta\phi}l_{\text{adv}}(\theta, \phi)[\nabla^2_{\phi\phi}l_{\text{adv}}(\theta, \phi)]^{-1}\nabla_\phi l_{\text{defend}}(\theta, \phi)$ and $\bar{h}_{\text{d}}^t:=E[h_{\text{d}}^t|\theta^{(t)}, \phi^{(t+1)}]$.

\vpara{Assumptions for convergence analysis.}
We make the following assumptions for the convergence analysis.

\emph{Assumption 1 (Lipschitz continuity).} $\nabla l_{\text{adv}}(\theta, \phi)$, $\overline{\nabla} l_{\text{defend}}(\theta, \phi)$, $\nabla \mathcal{L}_{\text{defend}}(\theta)$, $\phi^*(\theta)$ and $\nabla \phi^*(\theta)$ are Lipschitz continuous with constant $L_1$, $L_2$, $L_3$, $L_4$ and $L_5$.
\begin{align*}
    \| \nabla l_{\text{adv}}(\theta_1, \phi_1)-\nabla l_{\text{adv}}(\theta_2, \phi_2)\|&\leq L_1\|(\theta_1, \phi_1)-(\theta_2, \phi_2)\|, \\
    \| \overline{\nabla} l_{\text{defend}}(\theta, \phi^*(\theta))-\overline{\nabla} l_{\text{defend}}(\theta, \phi) \|&\leq L_2\|\phi^*(\theta)-\phi\|, \\
    \| \nabla \mathcal{L}_{\text{defend}}(\theta_1)- \nabla \mathcal{L}_{\text{defend}}(\theta_2) \|&\leq L_3\|\theta_1-\theta_2\|, \\
    \| \phi^*(\theta_1)-\phi^*(\theta_2) \|&\leq L_4\|\theta_1-\theta_2\|, \\
    \| \nabla\phi^*(\theta_1)-\nabla\phi^*(\theta_2) \|&\leq L_5\|\theta_1-\theta_2\|.
\end{align*}

\emph{Assumption 2 (Strong convexity).} For any fixed $\theta$, $l_{\text{adv}}(\theta, \phi)$ is $\mu$-strongly convex in $\phi$.

\emph{Assumption 3 (Bounded variance of stochastic gradients).} The variances of stochastic derivatives of $\nabla l_{\text{defend}}(\theta, \phi)$, $\nabla l_{\text{adv}}(\theta, \phi)$, $\nabla^2l_{\text{adv}}(\theta, \phi)$ are bounded by constant $\sigma^2_1$, $\sigma^2_2$, $\sigma^2_3$, respectively.

\emph{Assumption 4 (Bounded gradient estimates).} The stochastic gradient estimates satisfy: $E[\|h_{\text{d}}^t\|^2|\theta^{(t)}, \phi^{(t+1)}]\leq \tilde{C}^2$, $\|\bar{h}_{\text{d}}^t-\overline{\nabla}l_{\text{defend}}(\theta^{(t)}, \phi^{(t+1)})\|\leq b_t$, $E[\|h_{\text{d}}^t-\bar{h}_{\text{d}}^t\|^2]\leq \tilde{\sigma}^2$.

\vpara{Restatement of Theorem \ref{theorem:convergence}}.
\setcounter{theorem}{2} 
\begin{theorem} \label{theorem:convergece_detail}
Suppose Assumptions 1-4 hold. Let $\alpha_t$  and $\beta_t$ be the stepsizes when training the target model and the attack model, respectively. Choose $\alpha_t$ and $\beta_t$ as
\begin{equation*}
    \alpha_t=\text{min} \{\bar{\alpha}_1, \bar{\alpha}_2, \frac{\alpha}{\sqrt{T}} \}, \; \beta_t=\frac{8L_2L_4+2\eta L_5\tilde{C}^2\bar{\alpha}_1}{4T\mu}\alpha_t,
\end{equation*}
where $\alpha>0$ is a control constant. Constants $\bar{\alpha}_1$ and $\bar{\alpha}_2$ are defined as
\begin{equation*}
    \bar{\alpha}_1=\frac{1}{2L_3+4L_3L_4+\frac{2L_2L_5}{L_4\eta}}, \;
    \bar{\alpha}_2=\frac{8\frac{2\mu L_1}{\mu+L_1}}{(\mu+L_1)(8L_2L_4+2\eta L_5\tilde{C}^2\bar{\alpha}_1)},
\end{equation*}
where $\eta>0$ is a control constant. Then the outputs $\{\theta^{(t)}, \phi^{(t)}\}$ generated by the attack-defense arms race framework satisfy
\begin{equation*}
    \frac{1}{T}\sum_{t=1}^TE[\|\nabla \mathcal{L}_{\text{defend}}(\theta^{(t)})\|^2]=\mathcal{O}(\frac{1}{\sqrt{T}}) \; \text{and} \; E[\|\phi^{(T)}-\phi^*(\theta^{(T)})\|^2]=\mathcal{O}(\frac{1}{\sqrt{T}}). 
\end{equation*}
\end{theorem}

\vpara{Proof.}
The attack-defense arms race framework can be formulated as the following bilevel optimization problem:
\begin{align*}
&\text{minimize} \: \mathcal{L}_{\text{defend}}(\theta) := E_\xi[l_{\text{defend}}(\theta, \phi^*(\theta);\xi)] \\
& \text{subject to} \: \phi^*(\theta) = \argmin_\phi E_\zeta[l_{\text{adv}}(\theta, \phi;\zeta)],
\end{align*}
where $\xi$ and $\zeta$ are random variables associated stochastic gradient descent.
Then in the $(t+1)$-th iteration of the arms race framework, the model parameters $\theta$ and $\phi$ are updated as follows:
\begin{equation*}
    \phi^{(t+1)}=\phi^{(t)}-\beta_th_{\text{a}}^t, \; \theta^{(t+1)}=\theta^{(t)}-\alpha_th_{\text{d}}^t.
\end{equation*}
Let $\mathcal{T}_t^\prime$ denote the all the history information available up to iteration $t$, defined as $\mathcal{T}_t^\prime=\sigma\{\phi^1, \theta^1, \dots, \phi^{(t)}\}$, where $\sigma\{\cdot\}$ is the $\sigma$-algebra generated by the random variables.
Using the Lipschitz continuity of $\nabla \mathcal{L}_{\text{defend}}(\theta)$, we have
\begin{equation*}
    \mathcal{L}_{\text{defend}}(\theta^{(t+1)})\leq \mathcal{L}_{\text{defend}}(\theta^{(t)})+\langle \mathcal{L}_{\text{defend}}(\theta^{(t)}), \theta^{(t+1)}-\theta^{(t)}\rangle+\frac{L_3}{2}\|\theta^{(t+1)}-\theta^{(t)}\|^2.
\end{equation*}
Taking expectations, we have
\begin{align*}
    &E[\mathcal{L}_{\text{defend}}(\theta^{(t+1)})|\mathcal{T}_t^\prime]\\
    &\leq \mathcal{L}_{\text{defend}}(\theta^{(t)})+E[\langle\nabla \mathcal{L}_{\text{defend}}(\theta^{(t)}), \theta^{(t+1)}-\theta^{(t)}\rangle|\mathcal{T}_t^\prime]+\frac{L_3}{2}E[\|\theta^{(t+1)}-\theta^{(t)}\|^2|\mathcal{T}_t^\prime] \\
    &=\mathcal{L}_{\text{defend}}(\theta^{(t)}) -\alpha_t\langle\nabla \mathcal{L}_{\text{defend}}(\theta^{(t)}), \bar{h}_{\text{d}}^t\rangle+\frac{L_3\alpha_t^2}{2}E[\|h_{\text{d}}^t\|^2|\mathcal{T}_t^\prime] \\
    &=\mathcal{L}_{\text{defend}}(\theta^{(t)}) -\frac{\alpha_t}{2}\|\nabla \mathcal{L}_{\text{defend}}(\theta^{(t)})\|^2-\frac{\alpha_t}{2}\|\bar{h}_{\text{d}}^t\|^2+\frac{\alpha_t}{2}\|\nabla \mathcal{L}_{\text{defend}}(\theta^{(t)})-\bar{h}_{\text{d}}^t\|^2 +\frac{L_3\alpha_t^2}{2}\|\bar{h}_{\text{d}}^t\|^2+\frac{L_3\alpha_t^2}{2}E[\|h_{\text{d}}^t-\bar{h}_{\text{d}}^t\|^2|\mathcal{T}_t^\prime]\\
    &\leq \mathcal{L}_{\text{defend}}(\theta^{(t)}) -\frac{\alpha_t}{2}\|\nabla \mathcal{L}_{\text{defend}}(\theta^{(t)})\|^2-(\frac{\alpha_t}{2}-\frac{L_3\alpha_t^2}{2})\|\bar{h}_{\text{d}}^t\|^2+\frac{\alpha_t}{2}\|\nabla \mathcal{L}_{\text{defend}}(\theta^{(t)})-\bar{h}_{\text{d}}^t\|^2+\frac{L_3\alpha_t^2}{2}\tilde{\sigma}^2,
\end{align*}
where the first equation uses $\theta^{(t+1)}-\theta^{(t)}=-\alpha_th_{\text{d}}^t$ and $\bar{h}_{\text{d}}^t=E[h_{\text{d}}^t|\mathcal{T}_t^\prime]$, the second equation uses $\langle\nabla \mathcal{L}_{\text{defend}}(\theta^{(t)}), \bar{h}_{\text{d}}^t\rangle=\frac{1}{2}(\|\nabla \mathcal{L}_{\text{defend}}(\theta^{(t)})\|^2+\|\bar{h}_{\text{d}}^t\|^2-\|\nabla \mathcal{L}_{\text{defend}}(\theta^{(t)})-\bar{h}_{\text{d}}^t\|^2)$ and $\|h_{\text{d}}^t\|^2=\|\bar{h}_{\text{d}}^t+h_{\text{d}}^t-\bar{h}_{\text{d}}^t\|^2=\|\bar{h}_{\text{d}}^t\|^2+\|h_{\text{d}}^t-\bar{h}_{\text{d}}^t\|^2$, the last inequality uses $E[\|h_{\text{d}}^t-\bar{h}_{\text{d}}^t\|^2]\leq \tilde{\sigma}^2$.
Further, we have
\begin{align*}
    \|\nabla \mathcal{L}_{\text{defend}}(\theta^{(t)})-\bar{h}_{\text{d}}^t\|^2&=\|\overline{\nabla}l_{\text{defend}}(\theta^{(t)}, \phi^*(\theta^{(t)}))-\overline{\nabla}l_{\text{defend}}(\theta^{(t)}, \phi^{(t+1)})+\overline{\nabla}l_{\text{defend}}(\theta^{(t)}, \phi^{(t+1)})-\bar{h}_{\text{d}}^t\|^2\\
    &\leq 2\|\overline{\nabla}l_{\text{defend}}(\theta^{(t)}, \phi^*(\theta^{(t)}))-\overline{\nabla}l_{\text{defend}}(\theta^{(t)}, \phi^{(t+1)})\|^2+2\|\overline{\nabla}l_{\text{defend}}(\theta^{(t)}, \phi^{(t+1)})-\bar{h}_{\text{d}}^t\|^2 \\
    &\leq2L_2^2\|\phi^{(t+1)}-\phi^{(t)}\|^2+2b_t^2,
\end{align*}
where the last inequality uses Assumption 1 and Assumption 4. Taking expectations over $\mathcal{T}_t^\prime$ on both sides of the inequality, we have
\begin{align*}
    E[\mathcal{L}_{\text{defend}}(\theta^{(t+1)})]-E[\mathcal{L}_{\text{defend}}(\theta^{(t)})]\leq &-\frac{\alpha_t}{2}E[\|\nabla \mathcal{L}_{\text{defend}}(\theta^{(t)})\|^2]-(\frac{\alpha_t}{2}-\frac{L_3\alpha^2_t}{2})E[\|h_{\text{d}}^t\|^2]\\
    &+L_2^2\alpha_tE[\|\phi^{(t+1)}-\phi^*(\theta^{(t)})\|^2]+\alpha_t b_t^2+\frac{L_3\alpha^2_t}{2}\tilde{\sigma}^2.
\end{align*}
We then analyze the attack model optimization, decomposing the error of parameters of attack model as
\begin{equation*}
    \|\phi^{(t+1)}-\phi^*(\theta^{(t+1)})\|^2=\|\phi^{(t+1)}-\phi^*(\theta^{(t)})\|^2+\|\phi^*(\theta^{(t+1)})-\phi^*(\theta^{(t)})\|^2+2\langle\phi^{(t+1)}-\phi^*(\theta^{(t)}), \phi^*(\theta^{(t)})-\phi^*(\theta^{(t+1)})\rangle.
\end{equation*}
The upper bound of $\|\phi^{(t+1)}-\phi^*(\theta^{(t)})\|^2$ can be derived as
\begin{align*}
    E[\|\phi^{(t+1)}-\phi^*(\theta^{(t)})\|^2|\mathcal{T}_t^\prime]&=E[\|\phi^{(t)}-\beta_th_{\text{a}}^t-\phi^*(\theta^{(t)})\|^2|\mathcal{T}_t^\prime]\\
    &=\|\phi^{(t)}-\phi^*(\theta^{(t)})\|^2-2\beta_t\langle\phi^{(t)}-\phi^*(\theta^{(t)}), E[h_{\text{a}}^t|\mathcal{T}_t^\prime]\rangle+\beta_t^2E[\|h_{\text{a}}^t\|^2|\mathcal{T}_t^\prime] \\
    &\leq \|\phi^{(t)}-\phi^*(\theta^{(t)})\|^2-2\beta_t\langle\phi^{(t)}-\phi^*(\theta^{(t)}), \nabla l_{\text{adv}}(\theta^{(t)}, \phi^{(t)})\rangle+\beta_t^2\|\nabla l_{\text{adv}}(\theta^{(t)}, \phi^{(t)})\|^2+\beta_t^2\sigma_2^2 \\
    &\leq (1-\frac{2\mu L_1}{\mu+L_1}\beta_t)\|\phi^{(t)}-\phi^*(\theta^{(t)})\|^2+\beta_t(\beta_t-\frac{2}{\mu+L_1})\|\nabla l_{\text{adv}}(\theta^{(t)}, \phi^{(t)})\|^2+\beta_t^2\sigma_2^2\\
    &\leq (1-\frac{2\mu L_1}{\mu+L_1}\beta_t)\|\phi^{(t)}-\phi^*(\theta^{(t)})\|^2+\beta_t^2\sigma_2^2,
\end{align*}
where the first inequality uses $E[\|h_{\text{a}}^t\||\mathcal{T}_t^\prime]=\nabla l_{\text{adv}}(\theta^{(t)}, \phi^{(t)})$ and $\text{Var}[\|h_{\text{a}}^t\|]=E[\|h_{\text{a}}^t\|^2]-E[\|h_{\text{a}}^t\|]^2$
the second inequality uses the $\mu$-strong convexity and $L_1$ smoothness property of $l_{\text{adv}}(\theta, \phi)$, and the last inequality follows from choosing step size $\beta_t\leq\frac{2}{\mu+L_1}$. Taking expectations over $\mathcal{T}_t^\prime$ on both sides of the inequality, we have
\begin{equation*}
    E[\|\phi^{(t+1)}-\phi^*(\theta^{(t)})\|^2]\leq(1-\mu\beta_t)E[\|\phi^{(t)}-\phi^*(\theta^{(t)})\|^2]+\beta_t^2\sigma_2^2,
\end{equation*}
Using the smoothness of $\phi^*(\theta)$ and Assumption 4, we derive the upper bound of $\|\phi^*(\theta^{(t+1)})-\phi^*(\theta^{(t)})\|^2$:
\begin{align*}
    E[\|\phi^*(\theta^{(t+1)})-\phi^*(\theta^{(t)})\|^2]&\leq L_4^2E[\|\theta^{(t+1)}-\theta^{(t)}\|^2]\\
    &=L_4^2\alpha_t^2E[E[\|h_{\text{d}}^t-\bar{h}_{\text{d}}^t+\bar{h}_{\text{d}}^t\|^2|\mathcal{T}_t^\prime]]\\
    &\leq L_4^2\alpha_t^2(E[\|\bar{h}_{\text{d}}^t\|^2]+\tilde{\sigma}^2).
\end{align*}
The term $E[\langle\phi^{(t+1)}-\phi^*(\theta^{(t)}), \phi^*(\theta^{(t)})-\phi^*(\theta^{(t+1)})\rangle]$ can be decomposed into two terms:
\begin{align*}
    E[\langle\phi^{(t+1)}-\phi^*(\theta^{(t)}), \phi^*(\theta^{(t)})-\phi^*(\theta^{(t+1)})\rangle]=&-E[\langle\phi^{(t+1)}-\phi^*(\theta^{(t)}), \nabla \phi^*(\theta^{(t)})(\theta^{(t+1)}-\theta^{(t)})\rangle]\\
    &-E[\langle\phi^{(t+1)}-\phi^*(\theta^{(t)}), \phi^*(\theta^{(t+1)})-\phi^*(\theta^{(t)})-\nabla \phi^*(\theta^{(t)})(\theta^{(t+1)}-\theta^{(t)})\rangle].
\end{align*}
The first term is bounded by
\begin{align*}
    -E[\langle\phi^{(t+1)}-\phi^*(\theta^{(t)}), \nabla \phi^*(\theta^{(t)})(\theta^{(t+1)}-\theta^{(t)})\rangle]&=-\alpha_tE[\langle\phi^{(t+1)}-\phi^*(\theta^{(t)}), \nabla \phi^*(\theta^{(t)})\bar{h}_{\text{d}}^t\rangle]\\
    &\leq \alpha_tE[\|\phi^{(t+1)}-\phi^*(\theta^{(t)})\|\|\nabla \phi^*(\theta^{(t)})\bar{h}_{\text{d}}^t\|] \\
    &\leq \alpha_tL_4 E[\|\phi^{(t+1)}-\phi^*(\theta^{(t)})\|\|\bar{h}_{\text{d}}^t\|]\\
    &\leq 2\gamma_tE[\|\phi^{(t+1)}-\phi^*(\theta^{(t)})\|^2]+\frac{L_4^2\alpha_t^2}{8\gamma_t}E[\|\bar{h}_{\text{d}}^t\|^2],
\end{align*}
where the second inequality uses the Lipschitz continuity of $\phi^*(\theta)$, and the last inequality uses the Young's inequality that $ab\leq 2\gamma_ta^2+\frac{b^2}{8\gamma_t}$.
Using the smoothness of $\phi^*(\theta)$, the second term is bounded by
\begin{align*}
    &\quad -E[\langle\phi^{(t+1)}-\phi^*(\theta^{(t)}), \phi^*(\theta^{(t+1)})-\phi^*(\theta^{(t)})-\nabla \phi^*(\theta^{(t)})(\theta^{(t+1)}-\theta^{(t)})\rangle] \\
    &\leq E[\|\phi^{(t+1)}-\phi^*(\theta^{(t)})\|\|\phi^*(\theta^{(t+1)})-\phi^*(\theta^{(t)})-\nabla \phi^*(\theta^{(t)})(\theta^{(t+1)}-\theta^{(t)})\|]\\
    &\leq \frac{L_5}{2}E[\|\phi^{(t+1)}-\phi^*(\theta^{(t)})\|\|\theta^{(t+1)}-\theta^{(t)})\|^2]\\
    &\leq \frac{\eta L_5}{4}E[\|\phi^{(t+1)}-\phi^*(\theta^{(t)})\|^2E[\|\theta^{(t+1)}-\theta^{(t)})\|^2|\mathcal{T}_t^\prime]]+\frac{L_5}{4\eta}E[E[\|\theta^{(t+1)}-\theta^{(t)})\|^2|\mathcal{T}_t^\prime]]\\
    &\leq \frac{\eta L_5\tilde{C}^2\alpha_t^2}{4}E[\|\phi^{(t+1)}-\phi^*(\theta^{(t)})\|^2]+\frac{L_5\alpha_t^2}{4\eta}(E[\|\bar{h}_{\text{d}}^t\|^2]+\tilde{\sigma}^2),
\end{align*}
where the third inequality uses the Young's inequality that $1\leq\frac{\eta}{2}+\frac{1}{2\eta}$, and the last inequality uses Assumption 4.
Then we obtain the upper bound of $E[\langle\phi^{(t+1)}-\phi^*(\theta^{(t)}), \phi^*(\theta^{(t)})-\phi^*(\theta^{(t+1)})\rangle]$:
\begin{align*}
    E&[\langle\phi^{(t+1)}-\phi^*(\theta^{(t)}), \phi^*(\theta^{(t)})-\phi^*(\theta^{(t+1)})\rangle]\\
    &\leq(2\gamma_t+\frac{\eta L_5\tilde{C}^2}{4}\alpha_t^2)E[\|\phi^{(t+1)}-\phi^*(\theta^{(t)})\|^2]+(\frac{L_4^2\alpha_t^2}{8\gamma_t}+\frac{L_5\alpha_t^2}{4\eta})E[\|\bar{h}_{\text{d}}^t\|^2]+\frac{L_5\alpha_t^2}{4\eta}\tilde{\sigma}^2.
\end{align*}
Therefore, choosing $\gamma_t=L_2L_1\alpha_t$, the upper bound of the error of $\phi$ is
\begin{align*}
    E[\|\phi^{(t+1)}-\phi^*(\theta^{(t+1)})\|^2]&\leq(1+4L_2L_1\alpha_t+\frac{\eta L_5\tilde{C}^2}{2}\alpha_t^2)E[\|\phi^{(t)}-\phi^*(\theta^{(t)})\|^2]\\
    &+(L_4^2+\frac{L_4}{4L_2\alpha_t}+\frac{L_4\theta}{2\eta})\alpha_t^2E[\|h_{\text{d}}^t\|^2]+(L_4^2+\frac{L_5}{2\eta})\alpha_t^2\tilde{\sigma}^2.
\end{align*} 
Define the Lyapunov function as $\mathbb{V}^{(t)} :=\mathcal{L}_{\text{defend}}(\theta^{(t)})+\frac{L_2}{L_4}\|\phi^{(t)}-\phi^*(\theta^{(t)})\|$, then $E[\mathbb{V}^{(t+1)}]-E[\mathbb{V}^{(t)}]=E[\mathcal{L}_{\text{defend}}(\theta^{(t+1)})]-E[\mathcal{L}_{\text{defend}}(\theta^{(t)})]+\frac{L_2}{L_4}(E[\|\phi^{(t+1)}-\phi^*(\theta^{(t+1)})\|^2]-E[\|\phi^{(t)}-\phi^*(\theta^{(t)})\|^2])$.
Using the upper bounds of $E[\mathcal{L}_{\text{defend}}(\theta^{(t+1)})]-E[\mathcal{L}_{\text{defend}}(\theta^{(t)})]$ and $E[\|\phi^{(t+1)}-\phi^*(\theta^{(t+1)})\|^2]$, 
the difference between two Lyapunov functions is bounded as follows
\begin{align*}
    E[\mathbb{V}^{(t+1)}]-E[\mathbb{V}^{(t)}]\leq&-\frac{\alpha_t}{2}E[\|\nabla \mathcal{L}_{\text{defend}}(\theta^{(t)})\|^2]-(\frac{\alpha_t}{2}-\frac{L_3\alpha_t^2}{2}-L_2L_1\alpha_t^2-\frac{\alpha_t}{4}-\frac{L_2}{L_4}\frac{L_5}{2\eta})E[\|h_{\text{d}}^t\|^2]\\
    &+\frac{L_2}{L_4}((1+4L_2L_1\alpha_t+\frac{\eta L_5\tilde{C}^2}{2}\alpha_t^2)(1-\frac{2\mu L_1}{\mu+L_1}\beta_t)-1)E[\|\phi^{(t)}-\phi^*(\theta^{(t)})\|^2]\\
    &+\frac{L_2}{L_4}(1+5L_2L_1\alpha_t+\frac{\eta L_5\tilde{C}^2}{4}\alpha_t^2)\beta_t^2\sigma_2^2
    +\alpha_tb_t^2+(\frac{L_3}{2}+L_2L_1+\frac{L_5L_2}{2\eta L_1})\alpha_t^2\tilde{\sigma}^2.
\end{align*}
To ensure the decrease of $\mathbb{V}^{(t)}$, the following conditions must be met:
\begin{align*}
    \alpha_t&\leq\frac{1}{2L_3+4L_2L_4+\frac{2L_2L_5}{L_4\eta}},\\
    \frac{2\mu L_1}{\mu+L_1}\beta_t&\geq2L_2L_4\alpha_t+\frac{\eta L_5\tilde{C}^2}{2}\alpha_t^2, \\
    \beta_t&\leq\frac{2}{\mu+L_1}.
\end{align*}
To satisfy these conditions, we choose the stepsizes as defined in Theorem \ref{theorem:convergece_detail}, then the upper bound of the difference between the two Lyapunov functions is simplified as
\begin{align*}
    E[\mathbb{V}^{(t+1)}]-E[\mathbb{V}^{(t)}]
    &\leq-\frac{\alpha_t}{2}E[\|\nabla \mathcal{L}_{\text{defend}}(\theta^{(t)})\|^2]+c_1\alpha_t^2\sigma^2_2+\alpha_tb_t^2+c_2\alpha_t^2\tilde{\sigma}^2,
\end{align*}
where constants $c_1$ and $c_2$ are defined as
\begin{equation*}
    c_1=\frac{L_2}{L_1}(1+5L_2L_1\bar{\alpha}_1+\frac{\eta L_5\tilde{C}^2}{4}\bar{\alpha}_1^2)(\frac{8L_2L_1+2\eta L_5\tilde{C}^2\bar{\alpha}_1}{4\frac{2\mu L_1}{\mu+L_1}})^2,
\end{equation*}
\begin{equation*}
    c_2=(\frac{L_3}{2}+L_2L_1+\frac{L_5L_2}{2\eta L_4}).
\end{equation*}
Rewrite the inequality as $\frac{\alpha_t}{2}E[\|\nabla \mathcal{L}_{\text{defend}}(\theta^{(t)})\|^2]\leq E[\mathbb{V}^{(t)}]-E[\mathbb{V}^{(t+1)}]+c_1\alpha_t^2\sigma^2_2+\alpha_tb_t^2+c_2\alpha_t^2\tilde{\sigma}^2$ and apply telescoping from $t=1$ to $t=T$, we have
\begin{align*}
    \frac{1}{T}\sum_{t=1}^TE[\|\nabla \mathcal{L}_{\text{defend}}(\theta^{(t)})\|^2]&\leq\frac{\mathbb{V}^{(1)}+\sum_{t=1}^T(c_1\alpha_t^2\sigma^2_2+\alpha_tb_t^2+c_2\alpha_t^2\tilde{\sigma}^2)}{\frac{1}{2}\sum_{t=1}^T\alpha_t}\\
    &\leq\frac{2\mathbb{V}^{(1)}}{T\text{min}\{\bar{\alpha}_1,\bar{\alpha}_2\}}+\frac{2\mathbb{V}^{(1)}}{\alpha\sqrt{T}}+\frac{2c_1\alpha}{\sqrt{T}}\sigma^2_2+2b_t^2+\frac{2c_2\alpha}{\sqrt{T}}\tilde{\sigma}^2.
\end{align*}
Since $\frac{2\mathbb{V}^{(1)}}{T\text{min}\{\bar{\alpha}_1,\bar{\alpha}_2\}}$ is $\mathcal{O}(1/T)$, $\frac{2\mathbb{V}^{(1)}}{\alpha\sqrt{T}}$, $\frac{2c_1\alpha}{\sqrt{T}}\sigma^2_2$ and $\frac{2c_2\alpha}{\sqrt{T}}\tilde{\sigma}^2$ are $\mathcal{O}(1/\sqrt{T})$, we have $\frac{1}{T}\sum_{t=1}^TE[\|\nabla \mathcal{L}_{\text{defend}}(\theta^{(t)})\|^2]=\mathcal{O}(1/\sqrt{T})$.

\subsection{Proof of Theorem \ref{theorem: approx}}
\label{appendix: proof_approx}
Notably, the parameter change from $\theta_k$ to $\theta_k^\prime$ is related to the training loss of the shadow model.
We therefore examine how changes in the training dataset (i.e., $Z_k \rightarrow Z_k^\prime$) affect the training loss. Specifically, the parameters of the approximated shadow model on $D_k^{S^\prime}$ can be expressed as 
\begin{equation*} \label{eq:retrain}
\theta_k^\prime=\argmin _{\theta} \
L_{\mathcal{T}}(\theta;D_k^S)+\underset{z^\prime \in Z_k^\prime}{\sum}l(\theta;z^\prime) \
- \underset{z \in Z_k}{\sum}l(\theta;z),
\end{equation*}
where $L_{\mathcal{T}}$ is the loss function of the reference shadow model $f_k^{\text{ref}}$, and $l$ is the loss for each sample. 

Instead of training the shadow model from scratch, we approximate $\theta_k^\prime$ by introducing the influence of the perturbed samples $Z_k^\prime$ and removing the influence of the original samples $Z_k$, which is captured by a small parameter $\epsilon$:
\begin{equation*}
    \theta_{k, \epsilon}^\prime=\argmin _\theta L_{\mathcal{T}}(\theta;D^S)+\epsilon \underset{z^\prime \in Z_k^\prime}{\sum}l(\theta; z^\prime) - \epsilon \underset{z \in Z_k}{\sum}l(\theta; z).
\end{equation*}
To solve for $\theta_{k, \epsilon}^\prime$, we derive the optimality condition:
\begin{equation*}
    0=\nabla L_{\mathcal{T}}(\theta_{k, \epsilon}^\prime;D^S)+\epsilon \underset{z^\prime\in Z_k^\prime}{\sum} \nabla l(\theta_{k, \epsilon}^\prime; z^\prime)-\epsilon \underset{z \in Z_k}{\sum}\nabla l(\theta_{k, \epsilon}^\prime; z).
\end{equation*}
If $\epsilon$ is sufficiently small, then $\theta_{k, \epsilon}^\prime \rightarrow \theta_k$. Therefore, we can perform a first-order Taylor expansion at $\theta_k$ of the right-hand side:
\begin{equation*}
    0 \approx \nabla L_{\mathcal{T}}(\theta_k; D^S)+\epsilon \underset{z^\prime\in Z_k^\prime}{\sum} \nabla l(\theta_k; z^\prime)-\epsilon \underset{z \in Z}{\sum}\nabla l(\theta_k; z)+(\theta_{k, \epsilon}^\prime-\theta_k)[\nabla^2L(\theta_k; D_k^S)+\epsilon \underset{z^\prime\in Z_k^\prime}{\sum} \nabla^2 l(\theta_k; z^\prime)-\epsilon \underset{z \in Z}{\sum}\nabla ^2l(\theta_k; z)].
\end{equation*}
Since $\theta_k$ minimizes $L_{\mathcal{T}}(\theta; D_k^S)$, we have $\nabla L_{\mathcal{T}}(\theta_k; D^S)=0$. Solving for $\theta_{k, \epsilon}^\prime-\theta_k$, we get:
\begin{equation*}
    \theta_{k, \epsilon}’-\theta_k \approx -[\nabla^2 L_{\mathcal{T}}(\theta_k; D^S)+\epsilon \underset{z^\prime\in Z_k^\prime}{\sum} \nabla^2 l(\theta_k; z^\prime)-\epsilon \underset{z \in Z_k}{\sum}\nabla^2 l(\theta_k; z)]^{-1}[\epsilon \underset{z^\prime\in Z_k^\prime}{\sum} \nabla_\theta l(\theta_k; z^\prime)-\epsilon \underset{z \in Z_k}{\sum}\nabla l(\theta_k; z)],
\end{equation*}
where $\nabla^2 L_{\mathcal{T}}(\theta_k; D_k^S)$ is the Hessian $H_{\theta_k}$.
Dropping all terms in $\mathcal{O}(\epsilon)$, the equation simplifies to:
\begin{equation*}
    \theta_{k, \epsilon}^\prime-\theta_k \approx -H_{\theta_k}^{-1}[\underset{z^\prime\in Z_k^\prime}{\sum} \nabla_\theta l(\theta_k; z^\prime)-\epsilon \underset{z \in Z_k}{\sum}\nabla l(\theta_k; z)]\epsilon.
\end{equation*}
Since $\theta_k^\prime$ and $\theta_{k, \epsilon}^\prime$ are equivalent when $\epsilon = \frac{1}{|D_k^S|}$, we can linearly approximate the parameter change $\Delta(Z_k, Z_k^\prime)$ as
\begin{equation*}
    \Delta(Z_k, Z_k^\prime)=\frac{1}{|D_k^S|} \frac{d\theta_{k, \epsilon}’}{d\epsilon} \bigg|_{\epsilon=0}=-\frac{1}{|D_k^S|}H_{\theta_k}^{-1}[\underset{z^\prime\in Z_k^\prime}{\sum} \nabla_\theta l(\theta_k; z^\prime)-\epsilon \underset{z \in Z_k}{\sum}\nabla l(\theta_k; z)].
\end{equation*}

\clearpage
\newpage
\section{Complexity Analysis Details}
\label{appendix:complexity}
The time complexity of $\text{D}$-$\text{S}^2$HARE mainly involves three components: training reference shadow models, estimating parameters of approximated shadow models and executing the attack-defense arms race iteratively.
Suppose the number of shadow model and target model parameters is $p$, the size of the shadow model training data is $|D^S|$, the size of the target model training data is $|D|$. $T_s$ is the number of training iterations for target and shadow models, $T_a$ is the number of training iterations for the attack model, and $T$ is the number of iterations of the attack-defense arms race.
\begin{itemize}
    \item 
    \emph{Training reference shadow models}. 
    Suppose the input dimension of reference shadow model is $d$. For simplicity, suppose that reference shadow models are neural networks with $L$ hidden layers, each containing $l$ neurons. The output dimension of reference shadow models is $c$. The computational complexity of training a reference shadow model is linear in the number of model parameters, which is given by $\mathcal{O}((dl+l+(L-1)(l^2+l)+lc+c)|D^S|T_s)$, where $(dl+l+(L-1)(l^2+l)+lc+c)$ represents the number of model parameters $p$, $|D^S|$ is the size of training dataset and $T_s$ is the number of iterations. Therefore, the time complexity of training $K$ reference shadow models is $\mathcal{O}(Kp|D^S|T_s)$.
    \item 
    \emph{Estimating parameters of approximated shadow models}. We estimate the parameters of approximated shadow models using Eq.~\eqref{eq:est_para}, which involves calculating the inverse Hessian and the gradient terms. Following \citet{koh2017understanding}, the inverse Hessian can be approximated in $\mathcal{O}(p|B|)$ time, where $|B|$ denotes the number of training samples uniformly sampled from the dataset for approximation. Calculating the gradient terms requires $\mathcal{O}(p|Z|)$ time, where $|Z|$ represents the number of perturbed samples in the training dataset for the reference shadow model. Since $|Z|$ is typically small, the $\mathcal{O}(p|Z|)$ term can be ignored. Thus, the time complexity of estimating parameters for $N-K$ approximated shadow models simplifies to $\mathcal{O}((N-K)p|B|)$.
    \item 
    \emph{Executing the attack-defense arm-race}. The attack-defense arms race framework involves two main processes: Responsive CPI attack model refinement and target model update. Similar to the training of reference shadow models, training the Responsive CPI attack model over $N$ training samples requires $\mathcal{O}(pNT_a)$, and training the target model over a training dataset with $|D|$ samples requires $\mathcal{O}(p|D|T_s)$. These processes are executed over $T$ iterations, yielding a total complexity of $\mathcal{O}(T(pNT_a+p|D|T_s))$.
\end{itemize}
Summing the three components, the overall time complexity of $\text{D}$-$\text{S}^2$HARE is $\mathcal{O}(Kp|D^S|T_s+(N-K)p|B|+T(pNT_a+p|D|T_s))$. 

\clearpage
\newpage

\section{Implementation Details}
\label{appendix:implementation}

\vpara{Implementation details of trade-Off parameters.}
We provide additional details regarding the confidentiality-utility trade-off parameters used for various defense methods when evaluating their impact on the performance of confidentiality preservation and utility retention in Section~\ref{subsec:performance_recpi}. Specifically, for our method, the red points from left to right in the figure correspond to the performance under the trade-off hyperparameter \( \lambda \in \{0.1, 0.3, 0.5, 0.7\} \).
For Noisy Label, we flipped 10\%, 20\%, 30\%, 40\% of the labels for the target task, with the blue points in the figure arranged from right to left accordingly. In DP-SGD, we set the privacy budget $\epsilon$ used in its original paper~\citep{abadi2016deep} to the values in $\{0.1, 0.5, 1, 2\}$, corresponding to the green points from left to right in the figure. For Resampling, the purple points from right to left reflect down-sampling 20\%, 40\%, 60\%, 80\% of training samples with confidential property attribute $X_p=1$.
For Adversarial Defense, the orange points from right to left in the figure represent the performance under the
trade-off parameter between model optimization and defense
in its original paper~\citep{stockproperty}, 
with values in $\{0.2, 0.4, 0.6, 0.8\}$.

\vpara{Implementation of robustness checks.} 
When implementing unresponsive CPI attacks, we trained 500 shadow models. For logistic regression target model, we set the learning rate to 0.1 and L2 regularization coefficient to 1.0. The kernel width $\sigma^2$ was set to 0.015 when estimating training sample weight $r_i$. The trade-off parameters in robustness checks were kept the same as those in the main experiments, and all other implementation details followed the settings described in Section~\ref{subsec:setup}.

\clearpage
\newpage

\section{Sensitivity Analysis}
\label{appendix:sensitivity}
We conducted a sensitivity analysis to assess how the performance of our method is influenced by key hyperparameters: the number of reference shadow models $K$ used to simulate adversaries, and the perturbation budget $\delta$ used in the approximation strategy. For this analysis, we kept all other parameters consistent with the implementation details outlined in Section~\ref{subsec:setup}. We evaluated our method on the binary classification task of loan default rate.

\vpara{Number of reference shadow models.}
We varied the number of reference shadow models $K$ from 50 to 400, while keeping the total number of shadow models (reference and approximated) fixed at 500. The results can be found in Table \ref{tab:sensitivity_no_ref}.
When the number of reference shadow models is too small (e.g., $K=50$), the simulated adversary's ability to effectively infer the confidential property is significantly compromised. This limitation undermines our defense method's CPI-attack robustness, as it is not equipped with necessary defenses against adversaries with effective inference capabilities.
Conversely, when the number of reference shadow models is too large (e.g., $K=400$), our method's CPI-attack robustness improves. However, this enhancement comes at the cost of efficiency, as the runtime increases due to the need to train more reference shadow models from scratch.

\vspace{-2mm}
\begin{table}[H]
\centering
\caption{Sensitivity analysis of the number of reference shadow models.}
\label{tab:sensitivity_no_ref}
\renewcommand{\arraystretch}{0.5}
\resizebox{.75\textwidth}{!}
{
\begin{tabular}{@{} l | ccc | ccc @{}}
\Xhline{1pt}
\multicolumn{1}{c|}{{\begin{tabular}[c|]{@{}c@{}}  \end{tabular}}} &
 \multicolumn{3}{c |}{\begin{tabular}[c|]{@{}l@{}} White-box \end{tabular}} &
  \multicolumn{3}{c }{\begin{tabular}[c]{@{}l@{}} Black-box \end{tabular}}
  \\ & Attack success rate & Target accuracy & Run time  & Attack success rate &  Target accuracy & Run time   \\ \hline
$K=50$ & 0.60 & 0.898 & 119.24s & 0.53 & 0.897 & 225.32s\\
$K=100$ & 0.55 & 0.897 & 141.86s & 0.52 & 0.896 & 258.81s  \\
$K=200$ & 0.51 & 0.894 & 258.19s & 0.51 & 0.896 & 354.68s \\
$K=300$ & 0.51 & 0.894 & 374.67s & 0.50 & 0.895 & 458.84s \\
$K=400$ & 0.50 & 0.891 & 477.18s & 0.50 & 0.893 & 575.51s \\
\Xhline{1pt}
\end{tabular}}
\end{table}
\vspace{-5mm}

\vpara{Perturbation budget.} We also varied the perturbation budget $\delta$ used in our approximation strategy from 100 to 1500. Table \ref{tab:sensitivity_budget} shows the results. 
When the perturbation budget is too small, the approximated shadow models are nearly identical to the reference shadow models, limiting the effectiveness in simulating adversaries.
Conversely, when the perturbation budget is too large, the approximation error of the model parameters increases, also hindering the effectiveness of simulating adversaries.
This ineffectiveness in simulating adversaries, in turn, undermines our method's ability to defend against CPI attacks.
As a result, our model struggles to maintain confidentiality when the perturbation budget is too small or too large.
\vspace{-1cm}
\begin{table}[b]
\centering
\caption{Sensitivity analysis of perturbation budget.}
\label{tab:sensitivity_budget}
\renewcommand{\arraystretch}{0.5}
\resizebox{.6\textwidth}{!}
{
\begin{tabular}{l | cc | cc @{}}
\Xhline{1pt}
\multicolumn{1}{c|}{{\begin{tabular}[c|]{@{}c@{}}  \end{tabular}}} &
 \multicolumn{2}{c|}{\begin{tabular}[c|]{@{}l@{}} White-box \end{tabular}} &
  \multicolumn{2}{c }{\begin{tabular}[c]{@{}l@{}} Black-box \end{tabular}}
  \\ & Attack success rate & Target accuracy  & Attack success rate &  Target accuracy   \\ \hline
$\delta=100$ & 0.67 & 0.896 & 0.63 & 0.902 \\  
$\delta=200$ & 0.58 & 0.895 & 0.57 & 0.895 \\
$\delta=500$ & 0.50 & 0.892 & 0.50 & 0.890 \\
$\delta=1000$ & 0.55 & 0.897 & 0.52 & 0.896 \\
$\delta=1500$ & 0.66 & 0.898 & 0.60 & 0.900  \\
\Xhline{1pt}
\end{tabular}}
\end{table}

\clearpage
\newpage
\section{Notations}

\begin{table}[!h]
\centering
\caption{Summary of notations.}
\label{tab:notation}
\renewcommand{\arraystretch}{0.5}
{
\begin{tabular}{@{} c | l @{}}
\Xhline{1pt}
Notation & Description  \\ \hline
$D=(X,Y)$ & Target model training dataset with attribute matrix $X$ and target task label $Y$ \\
$X_p$ & Vector of the confidential property-related attribute $p$\\
$D^{\text{adv}}$ & Adversary's auxiliary dataset \\
$f_\theta$, $f_i^S$ & Target model with parameters $\theta$, shadow model \\
$\mathcal{F}_\theta$, $\mathcal{F}_i^S$ & Target model information, shadow model information \\
$P$, $P_i^S$ & Confidential property of target model, confidential property of shadow model $f_i^S$ \\
$h_\phi$ & Attack model with parameters $\phi$ \\
$\mathcal{D}_S$, $\mathcal{D}_T$ & Distribution of $\mathcal{F}_i^S$, $\mathcal{F}_\theta$ \\
$\mathcal{L}_\mathcal{A}$, $\mathcal{L}_\mathcal{P}$ & Training loss, test error of Responsive CPI attack model \\
$\mathcal{L}_\mathcal{T}$ & Loss function of the target model\\
$r_i$ & Training sample weight of Responsive CPI attack model \\
$\lambda$ & Trade-off parameter \\
$N$, $K$ & Number of all shadow models, number of reference shadow models \\
$f_k^{\text{ref}}$ & Reference shadow model \\
$D_k^S$, $D_k^{S^\prime}$ & Reference shadow model training dataset, perturbed dataset \\
$\theta_k$, $\theta_k^\prime$ & Reference shadow model parameters, approximated shadow model parameters \\
$Z_k$, $Z_k^\prime$ & Original set of samples in $D_k^S$, perturbed set of samples in $D_k^{S^\prime}$ \\
$\delta$ & Perturbation budget \\
\Xhline{1pt}
\end{tabular}}
\end{table}

\end{APPENDICES}

\end{document}